%% file: vis_in_connectomics_zib.tex
\pdfoutput=1

\newif\ifDraft

\documentclass[graybox]{svmult}

\usepackage{mathptmx}       
\usepackage{helvet}         
\usepackage{courier}        
\usepackage{type1cm}        
\usepackage{makeidx}         
\ifDraft
\usepackage[draft]{graphicx}
\else
\usepackage[final]{graphicx}
\fi
\usepackage{multicol}        
\usepackage[bottom]{footmisc}
\usepackage{url}
\usepackage{cite}
\usepackage{amssymb}
\usepackage{amsmath}

\usepackage{comment}

\usepackage{color} 

\makeindex             

\graphicspath{
{figures/}
}

\newlength{\mylength}

\title*{Visualization in Connectomics}
\titlerunning{Visualization in Connectomics}

\author{ Hanspeter Pfister, Verena Kaynig, Charl P.~Botha, Stefan Bruckner,\\ Vincent J.~Dercksen, Hans-Christian
Hege, and Jos B.T.M.~Roerdink}

\authorrunning{Pfister, Kaynig, Botha, Bruckner, Dercksen, Hege, Roerdink}

\institute{
Hanspeter Pfister and Verena Kaynig \at School of Engineering and Applied Sciences, Harvard University, 33 Oxford St., Cambridge, MA 02138,  \email{[pfister,vkaynig]@seas.harvard.edu}  \and
Charl P. Botha \at Computer Graphics and Visualization, Delft University of
Technology and Division of Image Processing, Leiden University Medical Center,
The Netherlands, \email{c.p.botha@tudelft.nl} \and
Stefan Bruckner \at Institute of Computer Graphics and Algorithms, Vienna University of Technology, Favoritenstra{\ss}e 9-11 / E186, A-1040 Wien, Austria, \email{bruckner@cg.tuwien.ac.at}
\and Vincent J. Dercksen and Hans-Christian Hege \at Department of Visualization and Data
Analysis, Zuse Institute Berlin, Takustra{\ss}e 7, D-14195 Berlin, Germany,
\email{[dercksen,hege]@zib.de}
\and Jos B.T.M. Roerdink \at Johann
Bernoulli Institute for Mathematics and Computer Science, University of Groningen, P.O. Box 407, 9700 AK Groningen, The
Netherlands, \email{j.b.t.m.roerdink@rug.nl}
}

\begin{document}

\maketitle
\textbf{Note:} Also to appear in the Dagstuhl 2012 SciVis book by
 Springer. Please cite this paper with its arXiv citation information.


\input{introduction}

\input{biologicalbackground}
\input{imagingmodalities}

\input{macroscale}

\input{mesoscale}

\input{microscale}

\input{integration}

\input{networkanalysis}
\input{conclusion}

\bibliographystyle{spmpsci}

\bibliography{vis_in_connectomics}

\end{document}

%% file: introduction.tex
\section{Introduction}
\label{sec:introduction}

Connectomics is a field of neuroscience that analyzes neuronal connections. A
connectome is a complete map of a neuronal system, comprising all neuronal connections
between its structures. The term `connectome' is close to the word `genome' and
implies completeness of all neuronal connections, in the same way as a genome
is a complete listing of all nucleotide sequences. The goal of connectomics
is to create a complete representation of the brain's wiring. Such a
representation is believed to increase our understanding of how functional brain
states emerge from their underlying anatomical structure~\cite{sporns:05}.
Furthermore, it can provide important information for the cure of neuronal dysfunctions like
schizophrenia or autism~\cite{seung:11}.

Different types of connectivity can be distinguished. \emph{Structural} or
anatomical connectivity usually refers to the ``wiring diagram'' of physical
connections between neural elements. These anatomical connections range in
scale from those of local circuits of single cells to large-scale networks of
interregional pathways~\cite{Sporns2010}.  \emph{Functional} connectivity is
defined as ``the temporal correlation between spatially remote
neurophysiological
events''~\cite{friston94:_funct_and_effec_connec_in_neuroim}. This can be seen
as a statistical property; it does not necessarily imply direct anatomical
connections. Finally, \emph{effective} connectivity concerns causal interactions
between distinct units within a nervous
system~\cite{friston94:_funct_and_effec_connec_in_neuroim}.

Sporns et al.~\cite{sporns:05} differentiate between macro-, meso- and
microscale connectomes. At the {\em macroscale} a whole brain can be imaged
and divided into anatomically distinct areas that maintain specific patterns
of interconnectivity. Spatial resolution at the macroscale is typically in the
range of millimeters. One order of magnitude smaller is the {\em mesoscale}
connectome that describes connectivity in the range of micrometers. At this
scale, local neuronal circuits, e.g., cortical columns, can be distinguished.
At the finest {\em microscale} the connectome involves mapping single neuronal
cells and their connectivity patterns.  Ultimately connectomes from all scales
should be merged into one hierarchical representation~\cite{sporns:05}. 

Independently of the scale, the connectivity can be represented as a
\emph{brain graph} $G(N;E)$ with nodes $N$ and weigthed edges $E$ representing
anatomical entities and the degree of structural or functional interactions,
respectively.  Associated to each abstract graph is a graph in real space that
connects real anatomical entities. Neural systems can be investigated by
analyzing topological and geometrical properties of these graphs and by
comparing them.  An equivalent way of representing an undirected or directed
brain graph is a \emph{connectivity} or \emph{association matrix} $C$, whose
entries $c_{ij}$ represent the degrees of interactions. Thresholding and
sometimes also binarizing them reveals the essential interactions. A spatial
connectivity graph can be depicted in real space, showing the actual physical
structure of the neural system. A connection matrix is usually visualized using
a color-encoding matrix view. For more details and examples see, e.g., the
recent reviews \cite{bullmore2011brain, Fornito2012}.

In contrast to genomics, the field of connectomics is to a large extent based
on image data. Therefore visualization of image data can directly support the 
analysis of brain structures and their structural or functional connections.

In this Chapter, we review the current state-of-the-art of visualization and
image processing techniques in the field of connectomics and describe some
remaining challenges. After presenting
some biological background in Section~\ref{sec:biological_background}
and an overview of relevant imaging modalities in
Section~\ref{sec:imaging_modalities}, we review current techniques to extract
connectivity information from image data at macro-, meso- and microscale in
Sections~\ref{sec:macroscale_connectivity}-\ref{sec:microscale_connectivity}.
Section~\ref{sec:integration} focuses on integration of anatomical
connectivity data.  The last section discusses visually supported analysis of
brain networks.

%% file: biologicalbackground.tex
\section{Biological Background}
\label{sec:biological_background}

\runinhead{Neural systems.} Functionally, neurons (or nerve cells) are the elementary signaling units of
the nervous system, including the brain.  Each neuron is composed of a cell
body (soma), multiple dendritic branches and one axonal tree, which receive
input from and transfer output towards other neurons, respectively. This
transfer is either chemical (synapses) or electrical (gap junctions).
Generally, during synaptic transmission vesicles containing neurotransmitter
molecules are released from terminals (boutons) on the axon of the presynaptic
neuron, diffuse across the synaptic cleft, and are bound by receptors on
dendritic spines of the postsynaptic neuron, inducing a voltage change, i.e.,\
a signal.

These basic building blocks can mediate complex behavior, as potentially large
numbers of them are interconnected to form local and long-range neural
microcircuits. At the meso-level, local neuron populations, e.g.,\ {\em cortical
minicolumns}, can be identified that act as elementary processing units. At the
macroscale, neurons in the human cortex are arranged in a number of
anatomically distinct areas, connected by interregional pathways called 
{\em tracts}~\cite{sporns:05}.

\runinhead{Model systems.} An important neuroscientific goal is to understand
how the human brain works.  However, due to its complexity (with an estimated
$10^{11}$ neurons with $10^{15}$ connections~\cite{sporns:05}), brain function
at the circuit or cellular level is often studied in other organisms that are more
amenable in complexity and size.

Conserved genes and pathways between different species offer the potential of
elucidating the mechanisms that affect complex human traits based on similar
processes in other organisms. This problem is particularly tractable in the
roundworm \emph{Caenorhabditis elegans}, whose brain with 302 neurons
has been completely mapped~\cite{White1986}, or in insects. In these organisms
brain structure and function can be studied at the level of single
identifiable neurons. Classical insect model organisms that are well
understood and allow easy genetic manipulations are fruit fly \emph{Drosophila
melanogaster} and the honeybee. Drosophila, for example, has been shown to be
an experimentally amenable model system even for the study of such
quintessential human physiological traits as alcoholism, drug abuse, or
sleep~\cite{Mackay2006}. 

Rodents, being mammals, have a brain structure that is similar but much smaller
than the human brain, and that therefore can be used to study cortical networks.
The mouse brain is an attractive model system to study, for example, the
visual system, due to the abundant availability of genetic tools allowing
monitoring and manipulating certain cell types or
circuits~\cite{Huberman2011}. The whisker-barrel pathway of the rat is a
relatively small and segregated circuit that is amenable to studying sensory
information processing at the molecular\slash synaptic, cell, and circuit\slash
region level.

%% file: imagingmodalities.tex
\section{Imaging Modalities Employed in Connectomics}
\label{sec:imaging_modalities}

\begin{figure}[htbp]
\centering
\vspace*{-10mm}
\includegraphics[width=0.8\columnwidth]{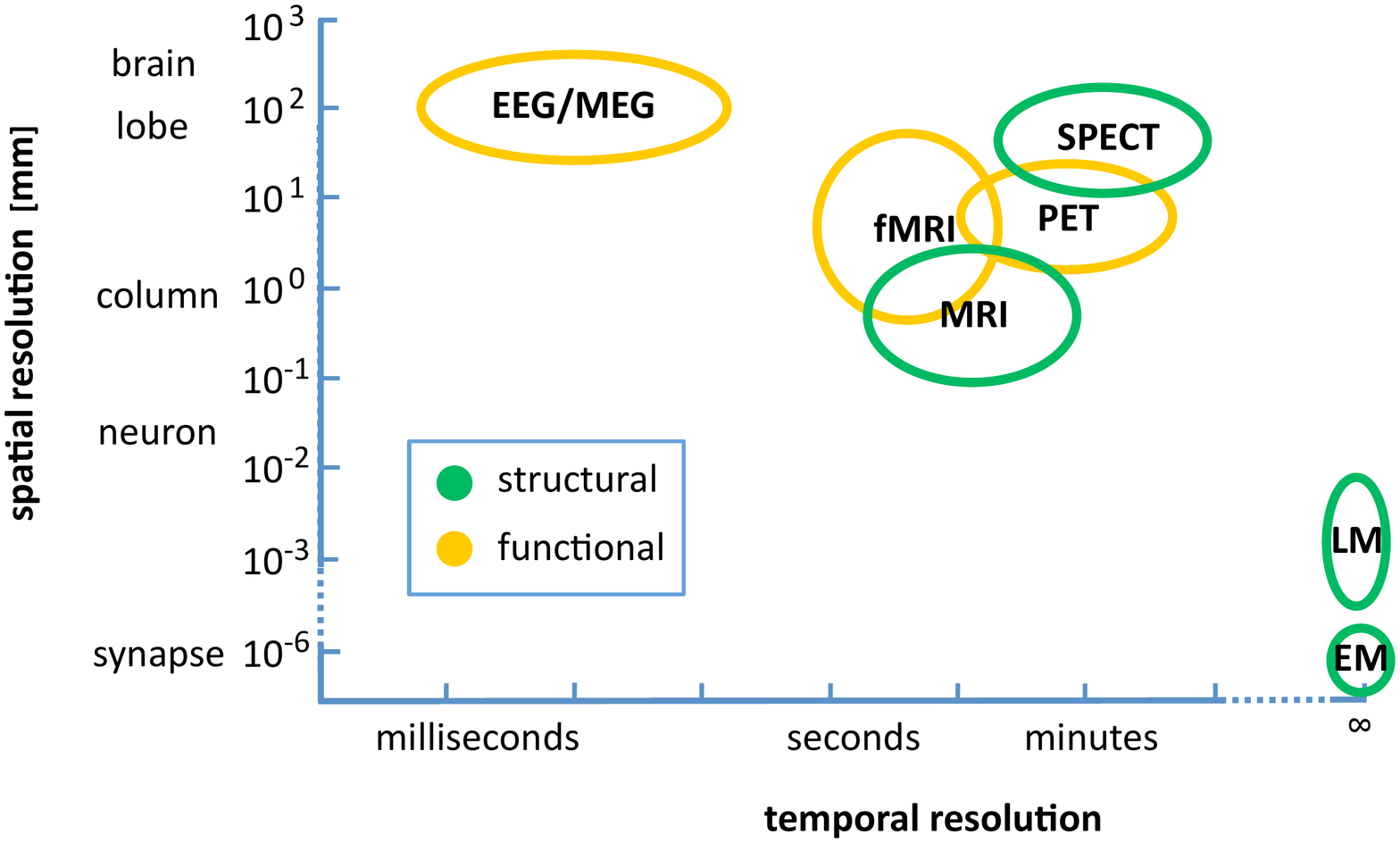}
\vspace*{-8mm}
\caption{Different brain imaging modalities and their spatial and temporal
resolutions. For connectomics, light- (LM) and electron microscopy (EM) are mostly
performed in vitro. The color indicates functional vs.\ structural information
in the acquired data.}
\label{figure:imageingModalities}
\end{figure}

We now provide an overview of imaging modalities that are used in
obtaining connectivity information. They differ in the
spatial and temporal resolution at which connectivity is captured.
At the \emph{macroscale} there is a wide range of structural and functional imaging
modalities, with applications in medical settings and anatomical research.
Functional imaging modalities include electroencephalography (EEG),
magnetoencephalography (MEG), functional magnetic resonance imaging (fMRI), 
and positron emission tomography (PET). Modalities like single-photon emission 
computed tomography (SPECT) and magnetic resonance imaging (MRI) provide 
structural information on the macroscale. Section \ref{sec:macroscale_connectivity}
gives a detailed introduction to the relevant modalities in the context of
connectomics. 
At the \emph{mesoscale} light microscopy (LM) techniques provide sufficient resolution
to image single neurons. Most light microscopy techniques focus on structural
imaging. Techniques like wide-field fluorescence microscopy allow for 
imaging of living cells, and computational optical sectioning microscopy 
techniques~\cite{Conchello2005} enable non-destructive acquisition of 3D data sets. 
Section \ref{sec:mesoscale_connectivity} provides further details about 
light microscopy techniques.
At the \emph{microscale} the sufficient resolution is offered by electron microscopy 
techniques (EM) such as Transmission Electron Microscopy (TEM) and Scanning Electron Microscopy (SEM).
These methods require technically complex specimen preparation and
are not applicable to live cell imaging. Imaging of 3D volumes requires
ultra-thin sectioning of the brain tissue followed by computational
realignment of the acquired images into one image volume \cite{kaynig:08}.
More information about electron microscopy in the connectomics setting
can be found in Section \ref{sec:microscale_connectivity}.
Figure~\ref{figure:imageingModalities} provides an overview of the different 
imaging modalities and their spatial and temporal resolution.
\vspace*{-4mm}



%% file: macroscale.tex
\section{Macroscale Connectivity}
\label{sec:macroscale_connectivity}

First, we discuss the main
acquisition techniques for revealing macroscopic functional and structural
connectivity. We start with MEG and EEG, as these were used for functional
connectivity before fMRI, then diffusion-weighted MRI for structural
connectivity, and finally fMRI for functional connectivity. Besides the
visualization approaches discussed here, the reader is also referred to
Section~\ref{sec:network_analysis} for more detail on network analysis and
comparative visualization techniques.

\subsection{EEG and MEG}

Developed in the 1920s, electroencephalography~(EEG) is the oldest
noninvasive functional neuroimaging technique, which records
electrical brain activity from electrodes on the scalp.  Nowadays, the
number of electrodes can be as large as 128 or even 512; in that case
one speaks of \emph{multichannel} or \emph{high-density}
EEG~\cite{schomer10:_nieder_elect}.  By contrast,
magnetoencephalography (MEG) measures magnetic fields outside the head
induced by electrical brain activity \cite{hamalainen93}. The temporal
frequency of these signals ranges from less than 1 Hz to over 100 Hz.
The spatial resolution is lower than for fMRI.  Sometimes, MEG is
preferred over EEG because the electrical signals measured by EEG
depend on the conduction through different tissues (e.g., skull and
skin).  However, EEG has much lower costs and higher equipment
transportability than MEG (and fMRI).  Moreover, EEG allows
participants more freedom to move than MEG and fMRI.  In
Section~\ref{sec:network_analysis} we will discuss the use of EEG to
discover functional brain networks.  Therefore, we will focus on EEG
for the remainder of this subsection.

Electrical potentials generated within the brain can be measured with
electrodes at the scalp during an EEG recording. The measured EEG
signals reflect rhythmical activity varying with brain state. Specific
brain responses can be elicited by the presentation of external
stimuli. For EEG analysis, one often studies activity in various
frequency bands, such as alpha, beta, theta or delta bands.  As a
result of \emph{volume conduction}, an electrical current flows from
the generator in the brain through different tissues (e.g., brain,
skull, skin) to a recording electrode on the scalp. The measured EEG
is mainly generated by neuronal (inhibitory and excitatory)
postsynaptic potentials and burst firing in the cerebral cortex.
Measured potentials depend on the source intensity, its distance from
the electrodes, and on the conductive properties of the tissues
between the source and the recording electrode.

Several visualization methods are applied to assist the interpretation
of the EEG.  In a conventional EEG visualization, the time-varying EEG
is represented by one time series per electrode, displaying the
measured potential as a function of time.  Synchronous activity
between brain regions is associated with a functional relationship
between those regions. EEG coherence, calculated between pairs of
electrode signals as a function of frequency, is a measure for this
synchrony.  A common visualization of EEG coherence is a graph layout.
In the case of EEG, graph vertices (drawn as dots) represent
electrodes and graph edges (drawn as lines between dots) represent
similarities between pairs of electrode signals.  Traditional visual
representations are, however, not tailored for multichannel EEG,
leading to cluttered representations. Solutions to this problem are
discussed in Section~\ref{sec:network_analysis}.

\subsection{MRI}

In magnetic resonance imaging, or MRI, unpaired protons, mostly in hydrogen
atoms, precess at a frequency related to the strength of the magnetic field
applied by the scanner. When a radio-frequency pulse with that specific
frequency is applied, the protons resonate, temporarily changing their
precession angle. They eventually regain their default precession angle, an
occurrence that is measured by the scanner as an electromagnetic signal. By
applying magnetic field gradients throughout three-dimensional space, protons at
different positions will precess and hence resonate at different frequencies,
enabling MRI to generate volume data describing the subject being scanned.

\subsubsection{Diffusion-Weighted Imaging}

Water molecules at any temperature above absolute zero undergo Brownian motion
or molecular diffusion~\cite{einstein_investigations_1905}. In free water,
this motion is completely random, and water molecules move with equal
probability in all directions. In the presence of constraining structures such
as the axons connecting neurons together, water molecules move more often in
the same direction than they do across these structures. When such a molecule
moves, the two precessing protons its hydrogen nucleus contains move as well.
When this motion occurs in the same direction as the diffusion gradient $q$
(an extra magnetic field gradient that is applied during scanning) of a
diffusion-weighted MRI scan, the detected signal from that position is
weakened.  By applying diffusion gradients in a number of different
directions, a dataset can be built up showing the 3D water diffusion at all
points in the volume, which in turn is related to the directed structures
running through those points.

\begin{figure}
 \centering
\includegraphics[height=0.42\textwidth]{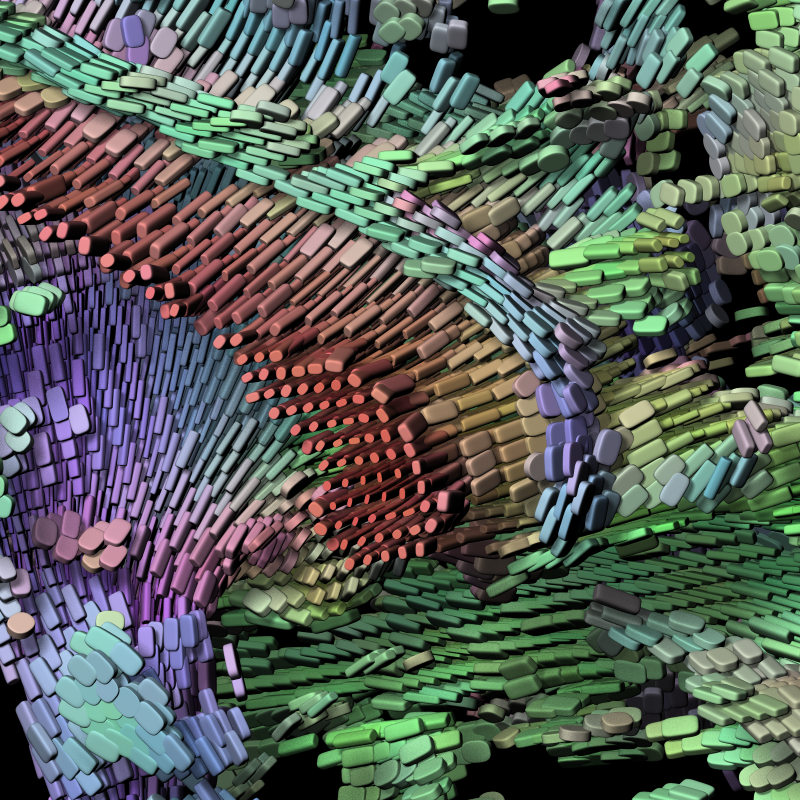}
\hspace*{1mm}
\includegraphics[height=0.42\textwidth]{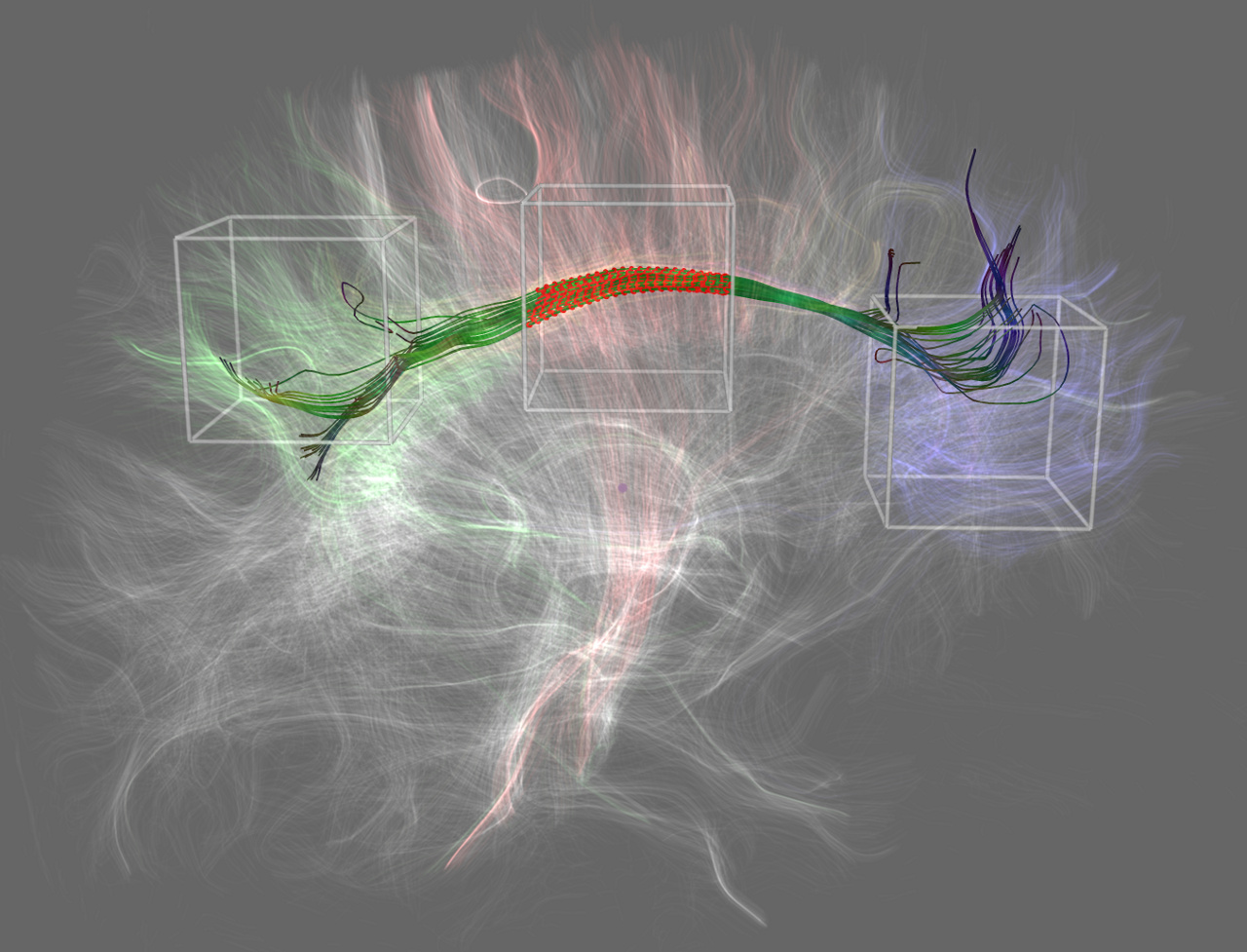}
\caption{On the left, superquadric glyphs have been used to represent
the diffusion tensors in a 3-D region of a brain
dataset~\cite{kindlmann_superquadric_2004}. On the right, the cingulum
neural fiber bundle has been highlighted in a full-brain
tractography~\cite{blaas_fast_2005}.}
\label{fig:dti}
\end{figure}

\runinhead{Diffusion tensor imaging.}
When at least six directions are acquired, a $3 \times 3$ symmetric diffusion
tensor can be derived, in which case the modality is described as Diffusion
Tensor Imaging (DTI). Per voxel DTI, often visualized with an ellipsoid, is
not able to represent more than one major diffusion direction through a voxel.
If two or more neural fibers were to cross, normal single tensor DTI would
show either planar or more spherical diffusion at that point. The left
image of Figure~\ref{fig:dti} shows a 3-D subset of such a dataset,
where each tensor has been represented with a superquadric glyph.

%
%
\noindent
DTI visualization techniques can be grouped into the following three classes~\cite{vilanova_introduction_2005}:

{\em Scalar metrics} reduce the multi-valued tensor data to one or
  more scalar values such as fractional anisotropy (FA), a measure of anisotropy
  based on the eigenvalues of the tensor, and then display the reduced data using
  traditional techniques, for example multi-planar reformation (slicing)
  or volume rendering. An often-used technique is to map the FA to intensity and
  the direction of the principal tensor eigenvector to color and then display
  these on a slice. Multiple anisotropy indices can also be used to define a
  transfer function for volume rendering, which is then able to represent
  the anisotropy and shape of the diffusion
  tensors~\cite{kindlmann_hue-balls_1999}.

{\em Glyphs} can be used to represent diffusion tensors without reducing
  the dimensionality of the tensor. In its simplest form, the eigensystem of the
  tensor is mapped directly to an ellipsoid. More information can be visually
  represented by mapping diffusion tensors to
  superquadrics~\cite{kindlmann_superquadric_2004}, see also
  Figure~\ref{fig:dti}.

{\em Vector- and tensor-field visualization} techniques
  visualize global information of the field. The best known is
  probably fiber tractography, where lines are reconstructed that
  follow the tensor data in some way and hence are related to the
  major directions of neural fibers. In its simplest form,
  streamlines, tangent to the principal eigenvectors of the diffusion
  tensors, are extracted and displayed~\cite{basser_vivo_2000}, where
  care has to be taken to terminate the streamlines in areas of
  isotropic or planar diffusion.  Hyperstreamlines take into account
  more of the tensor
  information~\cite{song_zhang_visualizing_2003}. Many tractography
  approaches require one or more regions of interest to be selected
  {\em before} tracts can be seeded starting only from those
  regions, while more recent efforts allow for full-brain fiber tracking
  followed by more intuitive interactive selection within the brain's
  tracked fiber bundles~\cite{Sherbondy2005,blaas_fast_2005}, see the
  right image in Figure~\ref{fig:dti} for an example.
  For a simplified visual representation, the envelopes of clustered
  streamline bundles can be shown~\cite{enders_visualization_2005}, or
  illustrative techniques such as depth-dependent halos can be
  used~\cite{Everts:2009:DDH}.  With probabilistic tractography,
  local probability density functions of diffusion or connectivity are
  estimated and can in turn be used to estimate the global
  connectivity, that is the probability that two points in the brain
  are structurally
  connected~\cite{behrens_characterization_2003}. This type of data is
  arguably a higher fidelity representation of structural
  connectivity.  Connectivity between two points can be visualized
  with, e.g., constant-probability isosurfaces, with direct
  volume rendering of the probability field, or using topological
  methods from flow
  visualization~\cite{schultz_topological_2007}. Calculating and
  effectively visualizing a full-brain probabilistic tractography
  would be challenging.


\runinhead{DSI and HARDI.}


As explained above, DTI is not able to capture more than one principal
direction per sample point. In order to reconstruct the full diffusion
probability density function (PDF), that is, the function describing the
probability of water diffusion from each voxel to all possible
displacements in the volume, about 500 or more diffusion-weighted MRI
volumes have to be acquired successively. This is called diffusion
spectrum imaging or DSI~\cite{hagmann_understanding_2006} and is the
canonical way of acquiring the complete 3-D water diffusion behavior.
However, the time and processing required to perform full DSI complicate
its use in research and practice.

In High Angular Resolution Diffusion Imaging, or HARDI, 
usually 40 or more directions are acquired in order to sample the 3-D diffusion
profile around every point~\cite{tuch_high_2002}. Based on such data, multiple
diffusion tensors can be fit to the data~\cite{tuch_high_2002}, higher order
tensors can be used~\cite{oezarslan_generalized_2003}, or a model-free method
such as Q-Ball imaging~\cite{tuch_qball_2004} can be applied. Q-Ball yields as
output an orientation distribution function, or ODF. The ODF is related to the
diffusion PDF in that it describes for each direction the sum of the PDF values
in that direction. It can be visualized as a deformed sphere whose radii
represents the amount of diffusion in the respective direction.

HARDI visualization follows much the same lines as DTI visualization, 
except that the data are more complex. Analogous to DTI, HARDI scalar
metrics, such as generalized (fractional) anisotropy and fractional
multifiber index, can be used to reduce the data to one or more scalar values
that can be visualized with traditional techniques.
Multiple diffusion tensors can be represented as glyphs, or the diffusion ODF
can be directly represented using a tesselated icosahedron or by raycasting
the spherical harmonics describing the ODF~\cite{peeters_fast_2009}. This
results in a field of complex glyphs representing at each point the diffusion
profile at that position. In contrast to DTI glyph techniques, regions of
crossing fibers can in general be identified.

Although there are fewer examples, especially in the visualization literature,
(probabilistic) fiber tracking can be performed based on HARDI
data~\cite{perrin_fiber_2005}. More recently, HARDI glyphs have been combined
dynamically with DTI glyphs and fiber tracts based on local data
characteristics~\cite{prckovska_fused_2011}.

\subsection{Functional MRI}
Blood-oxygen-level dependence, or BOLD, is a special type of MRI that is able
to measure increased levels of blood oxygenation~\cite{ogawa_brain_1990}. Due
to requiring more glucose from the bloodstream, active neurons cause higher
blood oxygenation in nearby veins. Based on this principle, functional MRI, or
fMRI, uses BOLD to image time-dependent 3-D neural activity in the
brain~\cite{ogawa_intrinsic_1992}.

fMRI can also be used to derive functional or {\em effective 
connectivity} in the brain. Functional connectivity is determined by calculating
the temporal correlations between the fMRI signals originating from different
parts of the brain~\cite{friston94:_funct_and_effec_connec_in_neuroim}. This is
done either whilst the subject performs a specific task, in order to assess how
the brain network is applied during that task, or during resting state, in order
to derive the baseline functional brain network. Connectivity data can be
determined between a specific seed region or voxel and one or more other
regions or voxels, or exhaustively between all regions or voxels in the brain.

Effective connectivity, defined as the causal influence one neuronal system exerts over
another, is dependent on a model of the connectivity between the participating
regions. For example, the signal at one position could be expressed as the
weighted sum of the signals
elsewhere~\cite{friston94:_funct_and_effec_connec_in_neuroim}. If the model is
invalid, the effective connectivity derived from fMRI is also invalid.

\begin{figure}
 \centering
\includegraphics[height=0.4\textwidth]{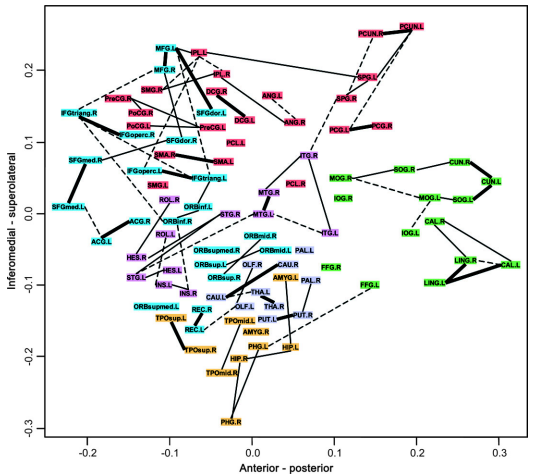}
\includegraphics[height=0.4\textwidth]{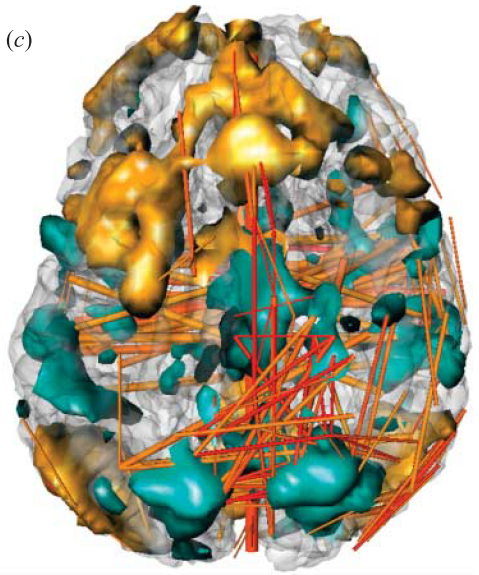}
\caption{On the left, different brain regions have been mapped to 2-D
using multi-dimensional scaling, with some of the strongest fMRI-derived
connections between them shown as
lines~\cite{salvador05cerebralcortex}. On the right, a number of the
strongest fMRI-derived connections between regions are shown in their
3-D anatomical context~\cite{worsley_comparing_2005}.}
\label{fig:fmri}
\end{figure}

Visualization of fMRI-derived connectivity information is quite varied,
often combining techniques from scientific and information
visualization. Scatter plots have been used to plot correlation strength
over distance, dendrograms and multi-dimensional scaling to represent
correlations between regions~\cite{salvador05cerebralcortex} (see
Figure~\ref{fig:fmri} left), matrix bitmaps to represent region-wise
correlation matrices~\cite{fair_maturing_2008}, 2-D and 3-D
(pseudo-)anatomical node-link diagrams to show the derived brain
networks~\cite{worsley_comparing_2005} (see Figure~\ref{fig:fmri}
right), and coupled-view visual analysis techniques to explore resting
state fMRI data~\cite{van_dixhoorn_visual_2010}. When connectivity is
determined between all pairs of voxels in the cortex, visualization and
knowledge extraction pose perceptual and computational challenges that
have not yet been fully explored.

%% file: mesoscale.tex

\section{Mesoscale Connectivity}
\label{sec:mesoscale_connectivity}

Light microscopy was the first modality that allowed for imaging of
single neuronal cells. While the resolution of a light microscope is
not sufficient to resolve synapses, it allows identifying major cell
parts, like dendrites, somas, axons, and also boutons as possible
locations for synaptic connections. Imaging whole neuronal cells and
analyzing their geometry enables neuroanatomists to identify different
types of cells and to come to conclusions about their function.
Following the motto ``the gain in the brain lies mainly in the
stain''~\cite{appel:97}, the three following main techniques
are employed to map neuronal circuits with light microscopy
\cite{lichtman:08}.

\runinhead{Single-cell staining by dye impregnation. } This is the
oldest staining method and it laid the foundation for modern
neuroscience. 
As neuronal tissue is
densely packed with cells, a complete staining of the whole sample
would not allow to discriminate single cells in light microscopy
images. Instead, the so-called {\em Golgi stain} enables stochastic 
marking of just a few individual nerve cells. The stained cells appear
dark in the light microscopy images, discriminating them from a bright
background formed by the unstained tissue. This staining method
combined with the ability of the light microscope to focus on
different depth of the sample allows for 3D imaging of
the cell geometry. The famous neuroscientist Cajal (1852-1934)
was able to identify different types of
neurons and also describe connectivity patterns and principles of
neuronal circuit organization using Golgi's method \cite{lichtman:08}.

\runinhead{Diffusion or transport staining. } 
Diffusion staining techniques enable biologists to analyze
the projective trajectory of brain regions. For this technique,
different staining markers are injected into different regions of the
brain \emph{in vivo}. The staining is then diffused along the connected
neurons. Finally a sample of brain tissue is extracted from a
different region, in which no marker has been injected. The color code
in the staining of different neurons in this area then reveals the
projection of these neurons back to the initial staining areas,
providing information about long-distance connectivity
\cite{gan:00}. The range of possible colors for this method
is limited to three or four different stainings. 

\begin{figure}[htbp]
\centering
\includegraphics[width=.6\columnwidth]{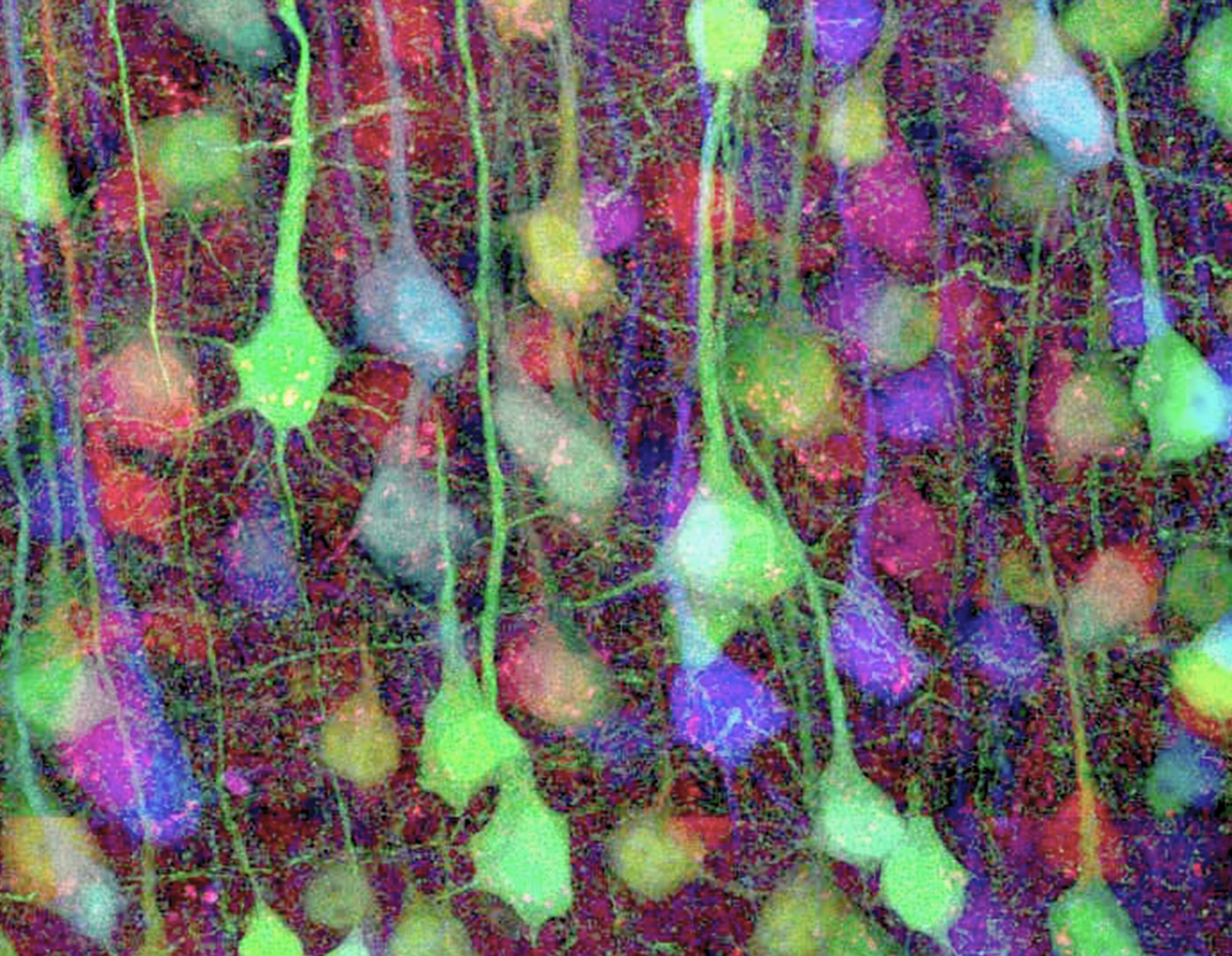}
\caption{Brainbow image of mouse cerebral cortex tissue. The different color stainings facilitate the differentiation of neuronal cells. Image curtesy of Jean Livet and Jeff Lichtman.}
\label{figure:brainbow}
\end{figure}

\runinhead{Multicolor or brainbow. } This staining technique does not
involve application or injection of staining to brain tissue. Instead,
transgenic mice are bred to produce photophysical fluorescent
proteins. A confocal laser-scanning microscope activates the
fluorescent proteins with a laser beam and records an image with the
expressed light. Brainbow mice are bred to express three fluorescent
proteins of different colors. By different stochastic expression of
these three colors, the single neurons of the mice are colored with
one out of $>100$ labels. The main advantage of this method is that it
allows to uniquely identify dendrites and axons belonging to the same
neuron in densely colored tissue~\cite{lichtman:08}, see also Figure \ref{figure:brainbow}. 
%

\bigskip

All of these three staining methods allow imaging the geometry of neurons at
the micrometer scale. The different staining protocols all aim at visually
separating single neurons out of the complex and dense neuronal tissue.
Visualization techniques for connectomics need to enhance the visual
separation further, e.g., by providing contrast enhancement and enabling
flexible mappings of image data to varying amounts of transparency in the
transfer function \cite{klein:09}. Especially for the brainbow staining it is
useful to have visual enhancement of color differences in regions of interest
where two neurons with a similar staining combination need to be
distinguished.  For diffusion staining this problem is less pronounced than
for brainbow data, as typically only three to four easily distinguishable
colors are used. But this also leads to the challenge of distinguishing two
neighboring cells that are stained with the same color. This problem also
arises in the Golgi stain, as only one color is applicable for this
staining. Thus visualization needs to focus on providing a good impression of
the neurons' geometry. The user needs to be able to access the
three-dimensional structure on different scale levels to infer the
connectivity of dendritic parts and axons.  In order to analyze the neuron
geometry further, dendritic and axonal trees have to be identified and
segmented. This task is typically performed either semi-automatically or fully
automatically with a final proof reading step \cite{turetken:11}.

An additional major challenge for the visualization of microscopy data sets in
the field of connectomics is the large data volume required to analyze the
geometry of full neurons. Microscopes typically only record regions of
interest at the required resolution. Afterwards the acquired images or image
stacks need to be stitched into one large data volume. While this problem is
well known and automatic methods for image stitching and alignment exist
\cite{preibisch:09, emmenlauer:09}, these tools typically work offline,
assembling all images into one large image file for later visualization. But
with image volumes in the gigapixel range this method is no longer
applicable. Instead, visualization tools are required to perform operations
 like image stitching, alignment, contrast enhancement, and
denoising {\em on-demand} in the region of interest shown to the user. To allow for
interactive visualization these operations do not only need to be executed
fast, but also on multiple scales, allowing the user to zoom in and out of the
displayed data volume. Recent work by Jeong et al. \cite{jeong:11}
provides this demand-driven visualization approach and combines it with a
client server architecture. The client can visualize the data with user interaction
and annotation while computations are performed on a high-performance server
transparently to the user. Multiple client instances can
connect to the same server to allow multiple users to access the data at the
same time and cooperatively work on the same data set.

%% file: microscale.tex

\section{Microscale Connectivity}
\label{sec:microscale_connectivity}

In contrast to light microscopy, which is limited in its resolution by the
wavelength of light, electron microscopy enables imaging neuronal tissue at the
nanometer scale. Hence, electron microscopy is the only imaging modality so
far that can resolve single synapses.  However, the sample preparation and image
acquisition in electron microscopy is labor-intensive and time-consuming.
As a consequence the analysis of the connectivity
between single neurons has been limited to sparse analysis of
statistical properties such as average synapse densities in different brain
regions \cite{dacosta:11}. Little is known about the complete connectivity
between single neurons. 
Information about the individual strength of synapses or the number of
connections between two cells can have important implications for
computational neuroanatomy and theoretical analysis of neuronal networks
\cite{valiant:06}.  

Recently, significant progress has been made in the automation of ultra-thin
serial sectioning \cite{hayworth:06} and automatic image acquisition
\cite{knott:08, denk:04}. These techniques allow neuroanatomists to acquire
large datasets of multiple terabytes (TB) in size. With a resolution of 5 nm per pixel,
and a section thickness of 50 nm, one cubic millimeter of brain tissue
requires imaging of 20,000 sections with 40 gigapixels per image, leading to an
image volume of 800 TB.  With data sets of this size new challenges emerge for
automatic computed analysis and visualization techniques. 
Important processing tasks include demand-driven image stitching 
and alignment, cell segmentation and 3D reconstruction, as well as multi-scale
visualization and multi-user interaction via client server architectures.

Electron microscopy samples are typically densely stained. While
in light microscopy sparse staining is necessary to visually separate a cell
of interest from unstained tissue background (see Section \ref{sec:mesoscale_connectivity}), 
the fine resolution of electron
microscopy allows to discriminate structures according to shape, size, and
texture. Electron microscopy images are limited to gray scale and 
typically do not have a uniform background. Instead,
the background is noisy and highly variable, which imposes an important challenge
for the visualization of electron microscopy image stacks. 
The image data cannot be
visualized according to gray values alone, as the densely stained tissue 
forms a nearly solid block. Instead, higher order features that discriminate
texture and shape, e.g., gradient histograms, are necessary to
enhance the visibility of different structures of interest in the
visualization \cite{jeong:10}. 
\begin{figure}[htbp]
 \centering
\includegraphics[width=\columnwidth]{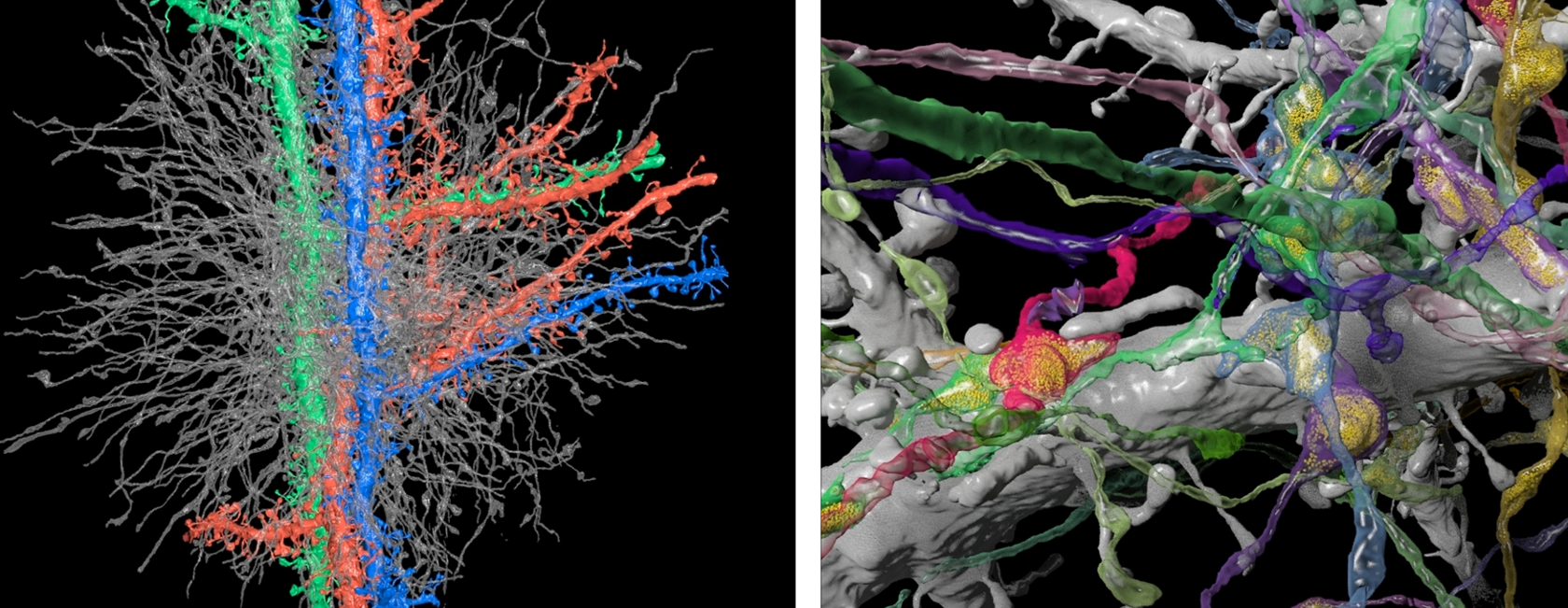}
\caption{Three dimensional reconstructions of neuronal structures from electron microscopy data. Left: three dendrites (colored) and all intervening axons (transparent), right: different axons (colored) with vesicle filled boutons (yellow).}
\label{figure:EM_reconstruction}
\end{figure}
Ultimately, full segmentation of the image data is necessary to allow the user
visual inspection of different biological structures, from small structures
such as vesicles or mitochondria to entire neuronal cells. Figure \ref{figure:EM_reconstruction} shows example reconstructions of different neuronal structures from electron microscopy images. 
A number of software packages have been developed to aid the user in manual segmentation of 
cell structures in the images~\cite{helmstaedter:11, cardona:10, fiala:05}. More recent semi-automatic
methods greatly facilitate this time-intensive process~\cite{roberts:11, straehle:11, chklovskii:10, vazquez:09}.

Progress has also been made on fully automatic segmentation of EM brain images
\cite{vazquez:11, jain:10, jurrus:10, kaynig:10a, kaynig:10b, vitaladevuni:10}. However, all methods developed so far require manual interaction and
inspection by users. Thus, visualization tools
should not only provide the ability to inspect the original EM data
and the computed segmentations, but also provide a user interface to detect and
correct segmentation errors, a process called {\em proofreading}.

Another interesting challenge for the visualization of neuronal microscopy
images is the concurrent display of light and electron microscopy data
acquired from the same sample. Correlative microscopy is a new developing
field, which allows inspection of the same neuronal tissue using both light
and electron microscopes. Thus the fine resolution of the electron microscopy images can be
combined with the advantage of color staining and information about long-range
connectivity in, e.g., diffusion stained light microscopy images.
Visualization of this data requires multi-modal registration of both data
sets, which has not yet been addressed for correlative microscopy.

Currently, most research efforts in connectomics at the microscale 
concentrate on the image acquisition and
segmentation of electron microscopy images. Little
research has been done in the visualization of entire connectomes, i.e.\ the
wiring diagram of neurons, their types and the connectivity for detailed
analysis of neuronal circuits. Connectomes, like the manually
reconstructed circuit of C.\ elegans, are visualized by connectivity matrices or
connection graphs~\cite{varshney:11}.

%% file: integration.tex
\section{Data Integration and Neural Network Modeling}
\label{sec:integration}

As described in the previous sections, neurobiological data can be acquired from many different sources. Relating these different kinds of data by integrating them in a common reference frame offers interesting opportunities to infer new knowledge about the relation between structure and function. In this section, we describe two approaches and their visualization aspects for such data integration with the purpose of inferring functional properties: Brain mapping and network modeling by reverse engineering.

\subsection{Brain mapping}
A major goal in neuroscience is to define the cellular architecture of the
brain. Mapping the fine anatomy of complex neuronal circuits is an
essential first step in investigating the neural mechanisms of information
processing. The term \emph{brain mapping} describes a set of neuroscience
techniques predicated on the mapping of biological quantities or properties
onto spatial representations of the brain resulting in maps. While all of
neuroimaging can be considered part of brain mapping, the term more
specifically refers to the generation of atlases, i.e., databases that combine
imaging data with additional information in order to infer functional
information. Such an undertaking relies on research and development in image
acquisition, representation, analysis, visualization, and interaction.
Intuitive and efficient visualization is important at all intermediate steps
in such projects. Proper visualization tools are indispensable for quality
control (e.g., identification of acquisition artifacts and
misclassifications), the sharing of generated resources among a network of
collaborators, or the setup and validation of an automated analysis pipeline.
Data acquired to study brain structure captures information on the brain at different scales (e.g., molecular, cellular,
circuitry, system, behavior), with different focus (e.g., anatomy, metabolism,
function), and  is multi-modal (text, graphics, 2D and 3D images, audio,
video)~\cite{Chicurel2000,Koslow2005}. The establishment of spatial relationships between initially unrelated
images and information is a fundamental step towards the exploitation of
available data~\cite{Bjaalie2002}. These relationships provide the basis for
the visual representation of a data collection and the generation of further
knowledge.

\runinhead{Databases and atlases. }
A neuroanatomical atlas serves as reference frame for comparing and integrating data from
different biological experiments. Maye et al.~\cite{Maye2006} give an
introduction and survey on the integration and visualization of neural
structures in brain atlases. Such atlases are an invaluable reference in
efforts to compile a comprehensive set of anatomical and functional data, and
in formulating hypotheses on the operation of specific neuronal circuits.

A classical image-based neuroanatomical atlas of Drosophila is the FlyBrain
atlas\footnote{\url{http://flybrain.neurobio.arizona.edu}}, spatially relating
a collection of 2D drawings, microscopic images, and text. One approach in
generating a digital atlas of this kind is by acquiring confocal microscope
images of a large number of individual brains. In each specimen, one or more
distinct neuronal types are highlighted using appropriate molecular genetic
techniques. Additionally, a general staining is applied to reveal the overall
structure of the brain, providing a reference for non-rigid registration to a
standard template. After registration, the specific neuronal types in each
specimen are segmented, annotated, and compiled into a database linked to the
physical structure of the brain. Jenett et al.~\cite{Jenett2006} describe
techniques for quantitative assessment, comparison, and
presentation of 3D confocal microscopy images of Drosophila brains and gene
expression patterns within these brains. Pereanu and
Hartenstein~\cite{Pereanu2006} and Rybak et al.~\cite{rybak2010digital},
described 3D atlases of the developing Drosophila brain and the honeybee brain.
The Neuroterrain 3D mouse brain atlas~\cite{Bertrand2008} consists of segmented
3D structures represented as geometry and references a large collection of
normalized 3D confocal images.

\runinhead{Visual exploration and analysis. }

\begin{figure}[tb]
  \centering
  \includegraphics[width=.49\textwidth]{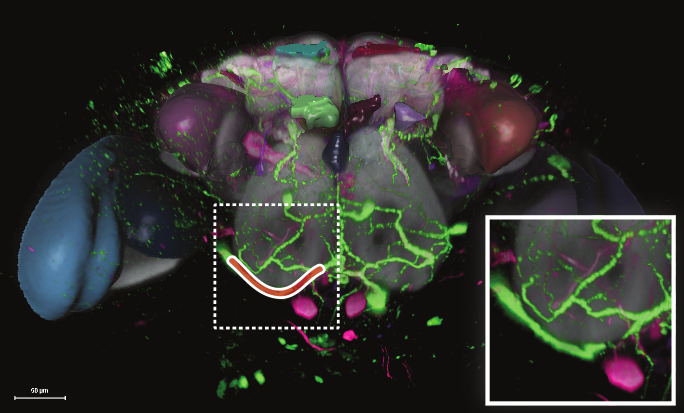}
  \includegraphics[width=.49\textwidth]{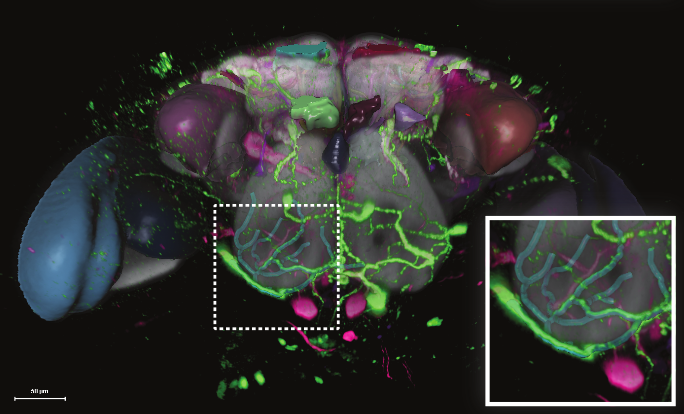}\\
  \caption{Visual query for neural projection in the Drosophila brain using the BrainGazer system~\cite{Bruckner2009a}.
  Left: The query is specified by sketching a path on top of a Gal4
  expression pattern. Right: An existing segmented neural projection
  that matches the query is displayed.}\label{fig:visualquery}
\end{figure}

3D microscopy data is often visualized using Maximum Intensity Projection
(MIP), which displays the maximum along viewing rays.
Direct Volume Rendering (DVR) enables better perception of spatial
relationships, but has the disadvantage of added complexity, as an additional
transfer function is required. It can lead to problems with occlusion
particularly when multiple channels need to be visualized simultaneously.
Maximum Intensity Difference Accumulation (MIDA)~\cite{Bruckner2009a} improves
this situation by combining the simplicity of MIP with additional spatial cues
provided by DVR. Wan et al.\cite{Wan2009} presented a tool for the
visualization of multi-channel data tailored to the needs of neurobiologists.
As acquired volumetric data is typically visualized together with segmented
structures, it is important to avoid occlusions as well as visual clutter.
Ku{\ss} et al.~\cite{Kuss2010} proposed and evaluated several techniques to
make spatial relationships more apparent.

However, to enable the exploration of large-scale collections of
neuroanatomical data, massive sets of data must be presented in a way that enables
them to be browsed, analyzed, queried and compared. An overview of a processing and visualization pipeline
for large collections of 3D microscopy images is provided in a study by de Leeuw et al.~\cite{deLeeuw2006}.
{NeuARt}~{II}~\cite{Burns2006} provides a general 2D visual interface to 3D
neuroanatomical atlases including interactive visual browsing by stereotactic
coordinate navigation. Brain Explorer~\cite{Lau2008}, an interface to the Allen Brain Atlas, allows the
visualization of mouse brain gene expression data in 3D. The CoCoMac-3D Viewer developed by Bezgin et
al.~\cite{Bezgin2009} implements a visual interface to two databases
containing morphology and connectivity data of the macaque brain for analysis
and quantification of connectivity data. An example of an interface to neuroanatomical image collections and databases
that features basic visual query functionalities is the European Computerized
Human Brain Database (ECHBD)~\cite{Fredriksson1999}. It connects a
conventional database with an infrastructure for direct queries on raster
data. Visual queries on image contents can be performed by interactive
definition of a volume of interest in a 3D reference image. Press et
al.~\cite{Press2001} focused on the graphical search within neuroanatomical
atlases. Their system called XANAT allows study, analysis, and storage of
neuroanatomical connections.  Users perform searches by graphically defining a
region of interest to display the connectivity information for this region.
Furthermore, their system supports also textual search using keywords
describing a particular region. Ku{\ss} et al.~\cite{Kuss2008} proposed
ontology-based high-level queries in a database of bee brain images based on
pre-generated 3D representations of atlas information. In the BrainGazer
system~\cite{Bruckner2009a} anatomical structures can be visually mined based
on their spatial location, neighborhood, and overlap with other structures. By
delineating staining patterns in a volume rendered image, for example, the
database can be searched for known anatomical objects in nearby locations (see
Figure~\ref{fig:visualquery}). Lin et al.~\cite{Lin2011} presented an approach
to explore neuronal structures forming pathways and 
circuits using connectivity queries. In order to explore the similarity and
differences of a large population of anatomical variations, Joshi et
al.~\cite{Joshi2009} proposed a similarity-space approach that embeds
individual shapes in a meta-space for content-driven navigation.

While these efforts represent promising directions, many challenges remain. As
noted by Walter et al.~\cite{Walter2010}, a major goal is the integration of
brain mapping data with other resources such as molecular sequences,
structures, pathways and regulatory networks, tissue physiology and
micromorphology. The ever-growing amount of data means that distributed
solutions are required. The integration of computational and human resources
gives significant benefits: each involved partner may bring computational
resources (in terms of hardware and tools), human resources (in terms of
expertise), and data to analyze. Advances in web technology, such as HTML5 and
WebGL, for instance,  provide new opportunities for visualization researchers
to make their work accessible to the neuroscience community.

\input{reverse_engineering}

%% file: reverse_engineering.tex
\subsection{Neural Network Modeling}

A complete reconstruction of the connectivity at synapse level is currently
possible for small brain volumes using electron microscopy techniques, but 
not yet feasible for volumes of the size of a cortical column.
Oberlaender et al.~\cite{Oberlaender2011} therefore pursue a reverse
engineering approach: A computational model of a cortical column in the rat somatosensory
cortex, consisting of $\sim$16,000  neurons, is created by integration of
anatomical data acquired by different imaging and reconstruction techniques
into a common reference system. As the data is acquired from different
animals in a population, the network represents an ``average'' cortical column: some
model parameters are given as probabilistic densities. By generating realizations 
of these stochastic parameters, concrete network models are created.

\begin{figure}[htbp] 
\centering
\includegraphics[width=\columnwidth]{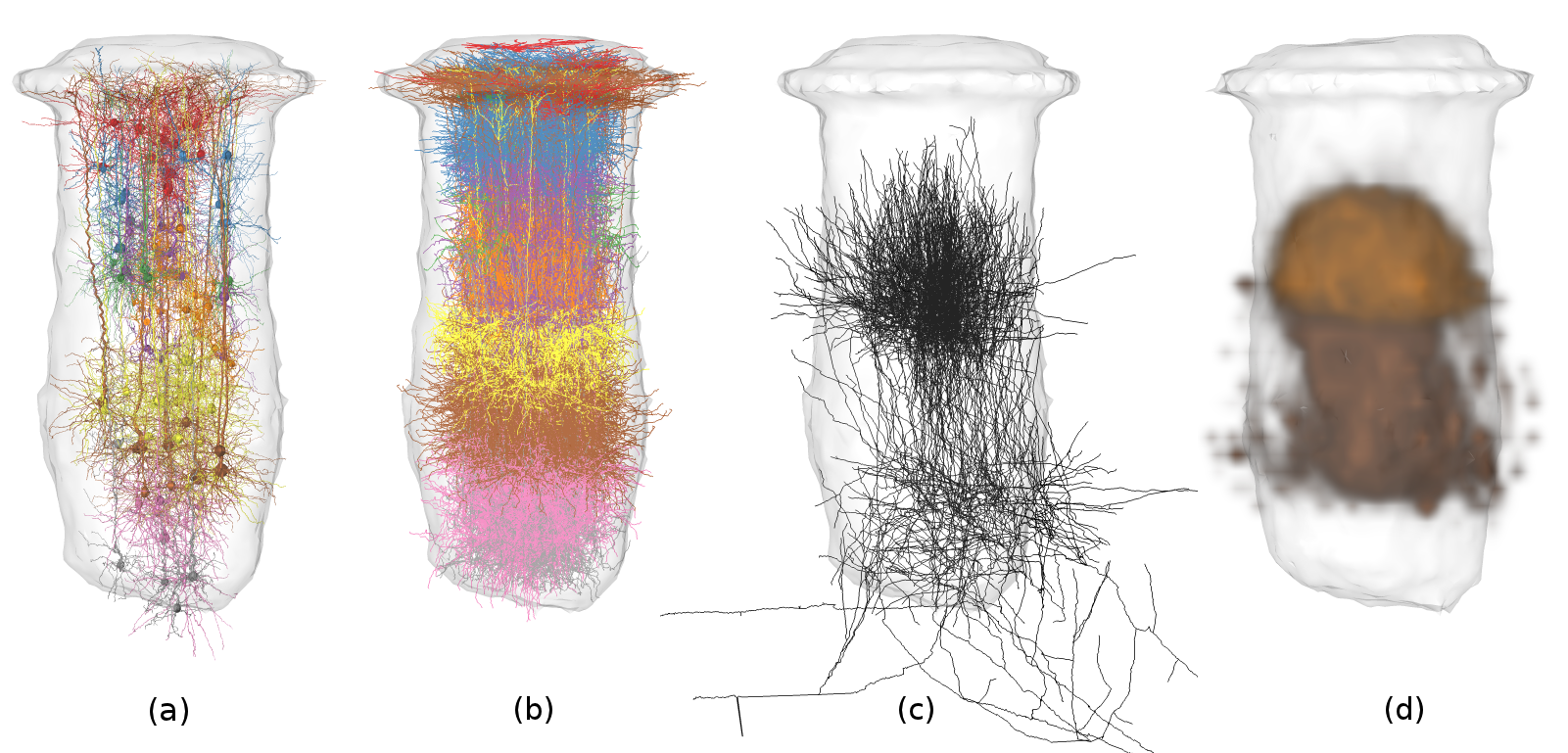} 
\caption{Reverse engineering of a cortical column. Reconstructed dendrites (a)
are replicated and inserted into the column reference frame according to a
given neuron density (b). By determining the local structural overlap with
axons projecting into the column (c) the number of synapses for different
post-synaptic cell types can be estimated. (d) Shown are synapse densities for
two cell types. Figure created from data published in~\cite{Oberlaender2011}.}
\label{figure:reverseengineering} 
\end{figure}

The number of neurons and their distribution in a cortical column is obtained
by automatic counting of neural soma (cell bodies) in confocal
images~\cite{Oberlaender:2009}. The 3D dendritic morphologies of $\sim$100
neurons of different cell types in the column as well as axons 
are reconstructed from transmitted light bright field
images~\cite{Dercksen2012}.  
The column model is created by generating soma positions satisfying the given
neuron density and replicating and inserting the dendrite morphologies into
the reference frame according to the given cell type frequency (see
Figure~\ref{figure:reverseengineering}). 
Differences in
synaptic densities between cell types can be quantified and
visualized~\cite{Oberlaender2011}. Based on the estimated number of
synapses per cell a complete network wiring is established to study network
function using numerical simulation~\cite{Lang2011}.

Extracting relevant neurobiological knowledge from such network models is a
challenging task. Whereas computation of specific quantities for comparison
with literature results in order to validate the model is straightforward,
exploratory knowledge discovery within such large, complex networks is not.
Easy-to-use tools are needed to let the neurobiologist query and visualize the
structural and functional properties of such networks or ensembles of network
realizations. As network models are increasing in size, large data handling
will be a challenging issue as well.

%% file: networkanalysis.tex

\section{Network Analysis and Comparative Visualization}
\label{sec:network_analysis}
A recent innovation in neuroimaging is connectivity analysis, in which
the anatomical or functional relation between different (underlying)
brain areas is calculated from data obtained by various modalities, allowing
researchers to study the resulting \emph{networks} of interrelated
brain regions. Of particular interest are \emph{comparisons} of
functional brain networks under different experimental conditions and 
between groups of subjects.

\subsection{Network Measures}

For each of the brain connectivity types (anatomical, functional, effective)
one can extract networks from data obtained by an appropriate brain imaging
modality \cite{bullmore04:_special_issue_funct_connec,koetter04:_onlin_cocom}.
The next step is to characterize such networks.  In the last decade a
multitude of \emph{topological} network measures have been developed in an
attempt to characterize and compare brain networks
\cite{Rubinov2009,stam07:_graph_theor_analy_of_compl,bullmore09:_compl,kaiser11}.
Such measures characterize aspects of global, regional, and local brain
connectivity\footnote{Similar approaches have been used in genomics
\protect\cite{sharan06:_model,luscombe04:_genom} and other areas.}. Examples of
global measures are characteristic path length, clustering coefficient,
modularity, centrality, degree distribution, etc. Some of them, such as
clustering coefficient or modularity, refer to \emph{functional segregation}
in the brain, i.e., the ability for specialized processing to occur in
densely interconnected groups of brain regions. Others characterize {\em functional
integration}, i.e., the ability to rapidly combine specialized information
from distributed brain regions
\cite{Rubinov2009,stam07:_graph_theor_analy_of_compl}. Typical measures in
this class are based on the concept of paths in the network, e.g.,
characteristic path length or global efficiency (average inverse shortest path
length). It is believed that both anatomical and functional brain connectivity
exhibit \emph{small-world properties}, i.e., they combine functionally segregated
modules with a robust number of intermodular links
\cite{bassett06:_small,sporns04}. The degree distribution can be used as a
measure of network resilience, i.e., the capacity of the network to
withstand network deterioration due to lesions or strokes.

For characterizing networks on a local scale one uses single node features
such as in-degree and out-degree, or the local clustering coefficient. Typical
regional network measures are \emph{network motifs}, which are defined as
patters of local connectivity. A typical motif in a directed network is a
triangle, consisting of feedforward and/or feedback loops. Both anatomical and
functional motifs are distinguished. The significance of a certain motif in a
network is determined by its frequency of occurrence, and the frequency of
occurrence of different motifs around a node is known as the motif fingerprint
of that node.

\subsection{Brain Network Comparison and Visualization}

The comparison of different brain networks presents challenging
problems.  Usually the networks differ in number and position of nodes
and links, and a direct comparison is therefore difficult.  One
possible approach is to compute a network measure for each of the
networks, and then compare the network measures.  However, this loses
spatial information.  For interpretation and diagnosis it may
be essential that local differences can be visualized in the original
network
representation~\cite{fair09:_funct_brain_networ_devel_from,shu09:_alter_anatom_networ_in_early}.
This asks for the development of mathematical methods, algorithms and
visualization tools for the \emph{local comparison} of complex
networks -- not necessarily of the same size -- obtained under
different conditions (time, frequency, scale) or pertaining to
different (groups of) subjects.

Several methods exist for spatial comparison of brain networks, which assume
that the position and number of network nodes is the same in the networks to
be compared. For example, Salvador et al.~\cite{salvador05cerebralcortex} use a brain parcellation based on a prior
standard anatomical template, dividing each cerebral hemisphere into 45
anatomical regions that correspond to the nodes of the brain network.  Another
possibility is to consider each voxel a network node, but in this way the
networks become very large. Links between the nodes can then be defined by
several measures of node-node association, such as correlation or mutual
information of temporal signals. Using the same construction for two or more
data sets enables a direct network comparison \cite{zalesky10:_networ}.

\begin{figure}[htbp] \centering
\includegraphics[width=0.7\columnwidth]{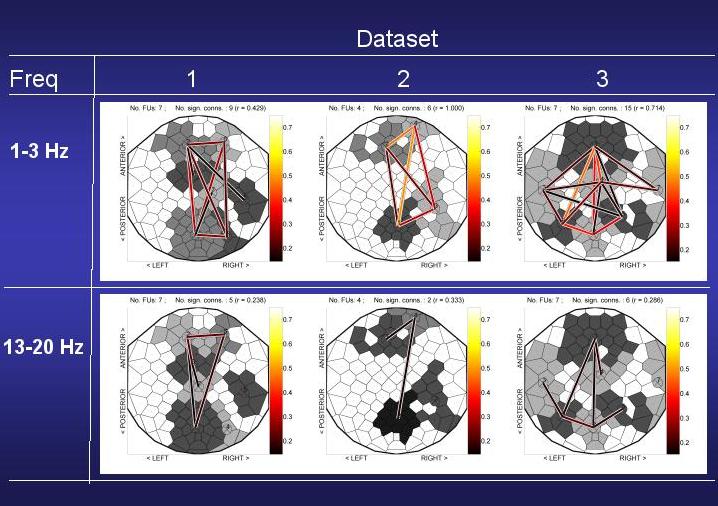} \caption{FU maps for
multichannel EEG coherence visualization. Brain responses were collected from
three subjects using an EEG cap with $119$ scalp electrodes. During a
so-called P$300$ experiment, each participant was instructed to count target
tones of $2000\,$Hz (probability~$0.15$), alternated with standard tones of
$1000\,$Hz (probability~$0.85$) which were to be ignored. After the
experiment, the participant had to report the number of perceived target
tones.  Shown are FU maps for target stimuli data, with FUs larger than 5
cells, for the 1-3Hz EEG frequency band (top row) and for 13-20Hz (bottom
row), for three datasets.} \label{figure:fumapslide} \end{figure}

A method to perform network comparison in the original network
representation was recently proposed for the case of multichannel EEG
by Crippa et al.~\cite{crippa11:_graph_eeg}.  This approach is
based on representation of an EEG coherence network by a so-called
\emph{functional unit} (FU), which is defined as a spatially connected
clique in the EEG graph, i.e., a set of electrodes used in the EEG
experiment that are spatially close and record pairwise significantly
coherent signals \cite{caat08:_data_driven_visual_group_analy}.  To
each electrode a Voronoi cell is associated and all cells belonging to
an FU are given a corresponding color. Lines connect FU centers if the
inter-FU coherence exceeds a significance threshold. The color of a
line depends on the inter-FU coherence.  Such a representation of the
FUs in an EEG recording is called a FU map. FU maps can be constructed
for different frequency bands or for different subjects (see
Figure~\ref{figure:fumapslide}).

Comparison of multiple FU maps can be done visually when displayed next to
each other, but this method is limited as humans are notoriously weak in
spotting visual differences in images. An alternative, which is more
quantitative although it still involves visual assessment to a certain
degree, is to compute a \emph{mean} FU map, based upon the concept of \emph{graph
averaging} \cite{crippa11:_graph_eeg}.  The mean of a set of input FU maps is
defined in such a way that it not only represents the mean group coherence
during a certain task or condition but also to some extent displays individual
variations in brain activity. The definition of a mean FU map relies on a
graph dissimilarity measure that takes into account both node positions and
node or edge attributes. A visualization of the mean FU map is used with a
visual representation of the frequency of occurrence of nodes and edges in the
input FUs. This makes it possible to investigate which brain regions are more
commonly involved in a certain task, by analyzing the occurrence of a FU of
the mean graph in the input FUs.

In \cite{crippa11:_graph_eeg} the graph averaging method was applied to the
analysis of EEG coherence networks in two case studies, one on mental fatigue
and one on patients with corticobasal ganglionic degeneration. An extension of
the method to resting state fMRI data was presented
in~\cite{crippa11:_data_driven_visual_funct_brain}.


%% file: conclusion.tex
\section{Conclusions}
\label{sec:conclusion}

There is currently great scientific interest in connectomics, as it is
believed to be an important prerequisite for understanding brain function. As
much of the data for obtaining neural connectivity is image-based, visualization
techniques are indispensable. Great effort has been put recently
into extraction of connectivity information from images, integration of
multimodal information into reference systems, and visual analysis of such data
and systems at different scales. These efforts will need to be intensified in
the future, as data is being produced at a much larger scale, also by new
image modalities. New methods to integrate this data across
modalities and scales to attain the ultimate goal, a description of the human 
connectome, will be the main challenges for visualization in connectomics.

%% file: vis_in_connectomics_zib.bbl
\begin{thebibliography}{100}
\providecommand{\url}[1]{{#1}}
\providecommand{\urlprefix}{URL }
\expandafter\ifx\csname urlstyle\endcsname\relax
  \providecommand{\doi}[1]{DOI~\discretionary{}{}{}#1}\else
  \providecommand{\doi}{DOI~\discretionary{}{}{}\begingroup
  \urlstyle{rm}\Url}\fi

\bibitem{appel:97}
Appel, N.M.: {Classical and Contemporary Histochemical Approaches for
  Evaluating Central Nervous System Microanatomy}.
\newblock Annals of the New York Academy of Sciences \textbf{820}(1), 14--28
  (1997)

\bibitem{basser_vivo_2000}
Basser, P.J., Pajevic, S., Pierpaoli, C., Duda, J., Aldroubi, A.: In vivo fiber
  tractography using {DT-MRI} data.
\newblock Magnetic Resonance in Medicine \textbf{44}(4), 625--632 (2000)

\bibitem{bassett06:_small}
Bassett, D.S., Bullmore, E.: Small-world brain networks.
\newblock Neuroscientist \textbf{12}(6), 512--523 (2006)

\bibitem{behrens_characterization_2003}
Behrens, T.E., Woolrich, M.W., Jenkinson, M., {Johansen‐Berg}, H., Nunes,
  R.G., Clare, S., Matthews, P.M., Brady, J.M., Smith, S.M.: Characterization
  and propagation of uncertainty in diffusion-weighted {MR} imaging.
\newblock Magnetic Resonance in Medicine \textbf{50}(5), 1077--1088 (2003)

\bibitem{Bertrand2008}
Bertrand, L., Nissanov, J.: {The Neuroterrain 3D Mouse Brain Atlas}.
\newblock Frontiers in neuroinformatics \textbf{2}, 3 (2008)

\bibitem{Bezgin2009}
Bezgin, G., Reid, A., Schubert, D., K\"{o}tter, R.: {Matching spatial with
  ontological brain regions using java tools for visualization, database
  access, and integrated data analysis}.
\newblock Neuroinformatics \textbf{7}(1), 7--22 (2009)

\bibitem{Bjaalie2002}
Bjaalie, J.G.: {Localization in the brain: New solutions emerging}.
\newblock Nature Reviews Neuroscience \textbf{3}, 322--325 (2002)

\bibitem{blaas_fast_2005}
Blaas, J., Botha, C.P., Peters, B., Vos, F.M., Post, F.H.: Fast and
  reproducible fiber bundle selection in {DTI} visualization.
\newblock In: C.~Silva, E.~Gr{\"o}ller, H.~Rushmeier (eds.) Proceedings of
  {IEEE} Visualization 2005, pp. 59--64 (2005)

\bibitem{Bruckner2009a}
Bruckner, S., Gr\"{o}ller, M.: {Instant Volume Visualization using Maximum
  Intensity Difference Accumulation}.
\newblock Computer Graphics Forum \textbf{28}(3), 775--782 (2009)

\bibitem{bullmore2011brain}
Bullmore, E., Bassett, D.: Brain graphs: graphical models of the human brain
  connectome.
\newblock Annual review of clinical psychology \textbf{7}, 113--140 (2011)

\bibitem{bullmore04:_special_issue_funct_connec}
Bullmore, E., Harrison, L., Lee, L., Mechelli, A., (eds.), K.F.: Special issue
  on functional connectivity.
\newblock Neuroinformatics \textbf{2}(2) (2004)

\bibitem{bullmore09:_compl}
Bullmore, E., Sporns, O.: Complex brain networks: graph theoretical analysis of
  structural and functional systems.
\newblock Nature Reviews Neuroscience \textbf{10}, 186--198 (2009).
\newblock See also Corrigendum (March 3, 2009)

\bibitem{Burns2006}
Burns, G.a.P.C., Cheng, W.C., Thompson, R.H., Swanson, L.W.: {The NeuARt II
  system: A viewing tool for neuroanatomical data based on published
  neuroanatomical atlases}.
\newblock BMC Bioinformatics \textbf{7}, 531--549 (2006)

\bibitem{caat08:_data_driven_visual_group_analy}
ten Caat, M., Maurits, N.M., Roerdink, J.B.T.M.: Data-driven visualization and
  group analysis of multichannel {EEG} coherence with functional units.
\newblock IEEE Trans.\ Visualization and Computer Graphics \textbf{14}(4),
  756--771 (2008)

\bibitem{cardona:10}
Cardona, A., Saalfeld, S., Preibisch, S., Schmid, B., Cheng, A., Pulokas, J.,
  Tomancak, P., Hartenstein, V.: An integrated micro- and macroarchitectural
  analysis of the drosophila brain by computer-assisted serial section electron
  microscopy.
\newblock PLoS Biol \textbf{8}(10), e1000,502 (2010)

\bibitem{Chicurel2000}
Chicurel, M.: {Databasing the brain}.
\newblock Nature \textbf{406}(6798), 822--825 (2000)

\bibitem{chklovskii:10}
Chklovskii, D.B., Vitaladevuni, S., Scheffer, L.K.: Semi-automated
  reconstruction of neural circuits using electron microscopy.
\newblock Current Opinion in Neurobiology \textbf{20}(5), 667 -- 675 (2010)

\bibitem{Conchello2005}
Conchello, J.A., Lichtman, J.: {Optical sectioning microscopy}.
\newblock Nature Methods \textbf{2}(12), 920--931 (2005)

\bibitem{crippa11:_graph_eeg}
Crippa, A., Maurits, N.M., Roerdink, J.B.T.M.: Graph averaging as a means to
  compare multichannel {EEG} coherence networks and its application to the
  study of mental fatigue and neurodegenerative disease.
\newblock Computers \& Graphics \textbf{35}(2), 265--274 (2011)

\bibitem{crippa11:_data_driven_visual_funct_brain}
Crippa, A., Roerdink, J.B.T.M.: Data-driven visualization of functional brain
  regions from resting state {fMRI} data.
\newblock In: P.~Eisert, K.~Polthier, J.~Hornegger (eds.) Proceedings Vision,
  Modeling and Visualization Workshop (VMV), 4-6 Oct, Berlin, pp. 247--254
  (2011)

\bibitem{dacosta:11}
Da~Costa, N.M., F\"{u}rsinger, D., Martin, K.A.C.: {The synaptic organization
  of the claustral projection to the cat's visual cortex.}
\newblock Journal of Neuroscience \textbf{30}(39), 13,166--13,170 (2010)

\bibitem{denk:04}
Denk, W., Horstmann, H.: {Serial Block-Face Scanning Electron Microscopy to
  Reconstruct Three-Dimensional Tissue Nanostructure}.
\newblock PLoS Biology \textbf{2}(11), e329 (2004)

\bibitem{Dercksen2012}
Dercksen, V.J., Oberlaender, M., Sakmann, B., Hege, H.C.: {Interactive
  Visualization -- a Key Prerequisite for Reconstruction of Anatomically
  Realistic Neural Networks}.
\newblock In: L.~Linsen (ed.) Visualization in Medicine and Life Sciences II.
  Springer (2012)

\bibitem{einstein_investigations_1905}
Einstein, A.: Investigations on the Theory of the Brownian Movement.
\newblock Dover (1956)

\bibitem{emmenlauer:09}
Emmenlauer, M., Ronneberger, O., Ponti, A., Schwarb, P., Griffa, A., Filippi,
  A., Nitschke, R., Driever, W., Burkhardt, H.: {XuvTools: free, fast and
  reliable stitching of large 3D datasets.}
\newblock Journal of Microscopy \textbf{233}(1), 42--60 (2009)

\bibitem{enders_visualization_2005}
Enders, F., Sauber, N., Merhof, D., Hastreiter, P., Nimsky, C., Stamminger, M.:
  Visualization of white matter tracts with wrapped streamlines.
\newblock In: {IEEE} Visualization, 2005. {VIS} 05, pp. 51-- 58. {IEEE} (2005)

\bibitem{Everts:2009:DDH}
Everts, M.H., Bekker, H., Roerdink, J.B.T.M., Isenberg, T.: Depth-dependent
  halos: illustrative rendering of dense line data.
\newblock IEEE Transactions on Visualization and Computer Graphics
  \textbf{15}(6), 1299--1306 (2009)

\bibitem{fair_maturing_2008}
Fair, D.A., Cohen, A.L., Dosenbach, N.U.F., Church, J.A., Miezin, F.M., Barch,
  D.M., Raichle, M.E., Petersen, S.E., Schlaggar, B.L.: The maturing
  architecture of the brain's default network.
\newblock Proceedings of the National Academy of Sciences \textbf{105}(10),
  4028 --4032 (2008)

\bibitem{fair09:_funct_brain_networ_devel_from}
Fair, D.A., Cohen, A.L., Power, J.D., Dosenbach, N.U.F., Church, J.A., Miezin,
  F.M., Schlaggar, B.L., Petersen, S.E.: Functional brain networks develop from
  a ``local to distributed'' organization.
\newblock PLoS Comput Biol \textbf{5}(5), e1000,381 (2009)

\bibitem{fiala:05}
Fiala, J.C.: Reconstruct: a free editor for serial section microscopy.
\newblock Journal of Microscopy \textbf{218}(1), 52--61 (2005)

\bibitem{Fornito2012}
Fornito, A., Zalesky, A., Pantelis, C., Bullmore, E.T.: Schizophrenia,
  neuroimaging and connectomics.
\newblock NeuroImage  (2012).
\newblock \doi{10.1016/j.neuroimage.2011.12.090}.
\newblock
  \urlprefix\url{http://www.sciencedirect.com/science/article/pii/S10538119120%
02133}.
\newblock (to appear)

\bibitem{Fredriksson1999}
Fredriksson, J.: {Design of an Internet accessible visual human brain database
  system}.
\newblock In: Proc. IEEE Int. Conf. on Multimedia Computing and Systems,
  vol.~1, pp. 469--474 (1999)

\bibitem{friston94:_funct_and_effec_connec_in_neuroim}
Friston, K.J.: Functional and effective connectivity in neuroimaging: {A}
  synthesis.
\newblock Human Brain Mapping \textbf{2}, 56--78 (1994)

\bibitem{gan:00}
Gan, W.B., Grutzendler, J., Wong, W.T., Wong, R.O.L., Lichtman, J.W.:
  {Multicolor ``DiOlistic'' Labeling Neurotechnique of the Nervous System
  Using}.
\newblock Neuron \textbf{27}, 219--225 (2000)

\bibitem{hagmann_understanding_2006}
Hagmann, P., Jonasson, L., Maeder, P., Thiran, J., Wedeen, V.J., Meuli, R.:
  Understanding diffusion {MR} imaging techniques: from scalar
  diffusion-weighted imaging to diffusion tensor imaging and beyond.
\newblock Radiographics \textbf{26 Suppl 1}, S205--S223 (2006)

\bibitem{hamalainen93}
H{\"a}m{\"a}l{\"a}inen, M., Hari, R., Ilmoniemi, R.J., Knuutila, J., Lounasmaa,
  O.V.: Magneto-encephalography -- theory, instrumentation, and applications to
  noninvasive studies of the working human brain.
\newblock Rev Mod Phys \textbf{65}, 413--497 (1993)

\bibitem{hayworth:06}
Hayworth, K.J., Kasthuri, N., Schalek, R., Lichtman, J.W.: {Automating the
  Collection of Ultrathin Serial Sections for Large Volume TEM
  Reconstructions}.
\newblock Microscopy and Microanalysis \textbf{12}(S02), 86--87 (2006)

\bibitem{helmstaedter:11}
Helmstaedter, M., Briggman, K.L., Denk, W.: High-accuracy neurite
  reconstruction for high-throughput neuroanatomy.
\newblock Nature Neuroscience \textbf{14}, 1081--1088 (2011)

\bibitem{Huberman2011}
Huberman, A.D., Niell, C.M.: {What can mice tell us about how vision works?}
\newblock Trends in Neurosciences \textbf{34}(9), 464--473 (2011)

\bibitem{jain:10}
Jain, V., Bollmann, B., Richardson, M., Berger, D.R., Helmstaedter, M.N.,
  Briggman, K.L., Denk, W., Bowden, J.B., Mendenhall, J.M., Abraham, W.C.,
  Harris, K.M., Kasthuri, N., Hayworth, K.J., Schalek, R., Tapia, J.C.,
  Lichtman, J.W., Seung, H.S.: Boundary learning by optimization with
  topological constraints.
\newblock IEEE Conference on Computer Vision and Pattern Recognition pp.
  2488--2495 (2010)

\bibitem{Jenett2006}
Jenett, A., Schindelin, J.E., Heisenberg, M.: {The Virtual Insect Brain
  protocol: Creating and comparing standardized neuroanatomy}.
\newblock BMC Bioinformatics \textbf{7}(1), 544--555 (2006)

\bibitem{jeong:10}
Jeong, W.K., Beyer, J., Hadwiger, M., Blue, R., Law, C., Vazquez-Reina, A.,
  Reid, R.C., Lichtman, J., Pfister, H.: Ssecrett and neurotrace: interactive
  visualization and analysis tools for large-scale neuroscience data sets.
\newblock IEEE Computer Graphics and Applications \textbf{30}(3), 58--70 (2010)

\bibitem{jeong:11}
Jeong, W.K., Johnson, M.K.: {Display-aware Image Editing}.
\newblock IEEE International Conference on Computational Photography  (2011)

\bibitem{Joshi2009}
Joshi, S.H., Horn, J.D.V., Toga, A.W.: {Interactive exploration of
  neuroanatomical meta-spaces}.
\newblock Frontiers in neuroinformatics \textbf{3}, 38 (2009)

\bibitem{jurrus:10}
Jurrus, E., Paiva, A.R.C., Watanabe, S., Anderson, J.R., Jones, B.W., Whitaker,
  R.T., Jorgensen, E.M., Marc, R.E., Tasdizen, T.: {Detection of neuron
  membranes in electron microscopy images using a serial neural network
  architecture.}
\newblock Medical Image Analysis \textbf{14}(6), 770--783 (2010)

\bibitem{kaiser11}
Kaiser, M.: A tutorial in connectome analysis: Topological and spatial features
  of brain networks.
\newblock NeuroImage \textbf{57}(3), 892 -- 907 (2011).
\newblock Special Issue: Educational Neuroscience

\bibitem{kaynig:08}
Kaynig, V., Fischer, B., Buhmann, J.M.: Probabilistic image registration and
  anomaly detection by nonlinear warping.
\newblock In: IEEE Conference on Computer Vision and Pattern Recognition, pp.
  1--8 (2008)

\bibitem{kaynig:10b}
Kaynig, V., Fuchs, T., Buhmann, J.M.: Geometrical consistent {3D} tracing of
  neuronal processes in {ssTEM} data.
\newblock In: International Conference on Medical Image Computing and Computer
  Assisted Intervention (2010)

\bibitem{kaynig:10a}
Kaynig, V., Fuchs, T., Buhmann, J.M.: Neuron geometry extraction by perceptual
  grouping in {ssTEM} images.
\newblock In: T.~Fuchs, J.M. Buhmann (eds.) IEEE Conference on Computer Vision
  and Pattern Recognition, pp. 2902--2909. IEEE (2010)

\bibitem{kindlmann_superquadric_2004}
Kindlmann, G.: Superquadric tensor glyphs.
\newblock In: Proceedings of {IEEE} {TVCG/EG} symposium on visualization, pp.
  147---154 (2004)

\bibitem{kindlmann_hue-balls_1999}
Kindlmann, G., Weinstein, D.: Hue-balls and lit-tensors for direct volume
  rendering of diffusion tensor fields.
\newblock In: Visualization '99. Proceedings, pp. 183--524. {IEEE} (1999)

\bibitem{klein:09}
Klein, J., Friman, O., Hadwiger, M., Preim, B., Ritter, F., Vilanova, A.,
  Zachmann, G., Bartz, D.: {Visual computing for medical diagnosis and
  treatment}.
\newblock Computers \& Graphics \textbf{33}(4), 554--565 (2009)

\bibitem{knott:08}
Knott, G., Marchman, H., Wall, D., Lich, B.: {Serial section scanning electron
  microscopy of adult brain tissue using focused ion beam milling.}
\newblock Journal of Neuroscience \textbf{28}(12), 2959--2964 (2008)

\bibitem{Koslow2005}
Koslow, S.H., Subramaniam, S.: {Databasing the Brain: From Data to Knowledge}.
\newblock Wiley (2005)

\bibitem{koetter04:_onlin_cocom}
K\"otter, R.: Online retrieval, processing, and visualization of primate
  connectivity data from the cocomac database.
\newblock Neuroinformatics \textbf{2}(2), 127--44 (2004)

\bibitem{Kuss2010}
Ku\ss, A., Gensel, M., Meyer, B., Dercksen, V., Prohaska, S.: {Effective
  Techniques to Visualize Filament-Surface Relationships}.
\newblock Computer Graphics Forum \textbf{29}(3), 1003--1012 (2010)

\bibitem{Kuss2008}
Ku{\ss}, A., Prohaska, S., Meyer, B., Rybak, J., Hege, H.C.: Ontology-based
  visualization of hierarchical neuroanatomical structures.
\newblock In: Proc. Eurographics Workshop on Visual Computing for Biomedicine,
  VCBM, Delft, The Netherlands, pp. 177--184 (2008)

\bibitem{Lang2011}
Lang, S., Dercksen, V.J., Sakmann, B., Oberlaender, M.: {Simulation of Signal
  Flow in Three-Dimensional Reconstructions of an Anatomically Realistic
  Neuronal Network in Rat Vibrissal Cortex}.
\newblock Neural Networks \textbf{24}(9), 998--1011 (2011)

\bibitem{Lau2008}
Lau, C., Ng, L., Thompson, C., Pathak, S., Kuan, L., Jones, A., Hawrylycz, M.:
  {Exploration and visualization of gene expression with neuroanatomy in the
  adult mouse brain}.
\newblock BMC Bioinformatics \textbf{9}, 153--163 (2008)

\bibitem{deLeeuw2006}
de~Leeuw, W., Verschure, P.J., van Liere, R.: {Visualization and analysis of
  large data collections: a case study applied to confocal microscopy data}.
\newblock IEEE Transactions on Visualization and Computer Graphics
  \textbf{12}(5), 1251--1258 (2006)

\bibitem{lichtman:08}
Lichtman, J., Livet, J., Sanes, J.: A technicolour approach to the connectome.
\newblock Nature Reviews Neuroscience \textbf{9}(6), 417--422 (2008)

\bibitem{Lin2011}
Lin, C.Y., Tsai, K.L., Wang, S.C., Hsieh, C.H., Chang, H.M., Chiang, A.S.: {The
  Neuron Navigator: Exploring the information pathway through the neural maze}.
\newblock In: Proceedings of IEEE Pacific Visualization 2011, pp. 35--42. IEEE
  (2011)

\bibitem{luscombe04:_genom}
Luscombe, N.M., Babu, M.M., Yu, H., Snyder, M., Teichmann, S.A., Gerstein, M.:
  Genomic analysis of regulatory network dynamics reveals large topological
  changes.
\newblock Nature \textbf{431}, 308--312 (2004)

\bibitem{Mackay2006}
Mackay, T.F., Anholt, R.R.: {Of flies and man: Drosophila as a model for human
  complex traits}.
\newblock Annual Review of Genomics and Human Genetics \textbf{7}, 339--367
  (2006)

\bibitem{Maye2006}
Maye, A., Wenckebach, T.H., Hege, H.C.: {Visualization, reconstruction, and
  integration of neuronal structures in digital brain atlases}.
\newblock The International journal of neuroscience \textbf{116}(4), 431--59
  (2006)

\bibitem{Oberlaender:2009}
Oberlaender, M., Dercksen, V.J., Egger, R., Gensel, M., Sakmann, B., Hege,
  H.C.: {Automated three-dimensional detection and counting of neuron somata}.
\newblock Journal of Neuroscience Methods \textbf{180}(1), 147--160 (2009)

\bibitem{Oberlaender2011}
Oberlaender, M., de~Kock, C.P.J., Bruno, R.M., Ramirez, A., Meyer, H.S.,
  Dercksen, V.J., Helmstaedter, M., Sakmann, B.: {Cell Type-Specific
  Three-Dimensional Structure of Thalamocortical Circuits in a Column of Rat
  Vibrissal Cortex}.
\newblock Cerebral Cortex  (2011).
\newblock \doi{doi: 10.1093/cercor/bhr317}

\bibitem{ogawa_brain_1990}
Ogawa, S., Lee, T.M., Kay, A.R., Tank, D.W.: Brain magnetic resonance imaging
  with contrast dependent on blood oxygenation.
\newblock Proceedings of the National Academy of Sciences \textbf{87}(24), 9868
  --9872 (1990)

\bibitem{ogawa_intrinsic_1992}
Ogawa, S., Tank, D.W., Menon, R., Ellermann, J.M., Kim, S.G., Merkle, H.,
  Ugurbil, K.: Intrinsic signal changes accompanying sensory stimulation:
  functional brain mapping with magnetic resonance imaging.
\newblock Proceedings of the National Academy of Sciences \textbf{89}(13),
  5951--5955 (1992)

\bibitem{oezarslan_generalized_2003}
\"Ozarslan, E., Mareci, T.H.: Generalized diffusion tensor imaging and
  analytical relationships between diffusion tensor imaging and high angular
  resolution diffusion imaging.
\newblock Magnetic Resonance in Medicine \textbf{50}(5), 955--965 (2003)

\bibitem{peeters_fast_2009}
Peeters, T.H., Pr\v{c}kovska, V., van Almsick, M., Vilanova, A., ter
  Haar~Romeny, B.M.: Fast and sleek glyph rendering for interactive {HARDI}
  data exploration.
\newblock In: Visualization Symposium, 2009. {PacificVis} '09. {IEEE} Pacific,
  pp. 153--160. {IEEE} (2009)

\bibitem{Pereanu2006}
Pereanu, W., Hartenstein, V.: {Neural Lineages of the Drosophila Brain: A
  Three-Dimensional Digital Atlas of the Pattern of Lineage Location and
  Projection at the Late Larval Stage}.
\newblock The Journal of Neuroscience \textbf{26}(20), 5534--5553 (2006)

\bibitem{perrin_fiber_2005}
Perrin, M., Poupon, C., Cointepas, Y., Rieul, B., Golestani, N., Pallier, C.,
  Rivière, D., Constantinesco, A., Bihan, D., Mangin, J.F.: Fiber tracking in
  {q-Ball} fields using regularized particle trajectories.
\newblock In: G.E. Christensen, M.~Sonka (eds.) Information Processing in
  Medical Imaging, vol. 3565, pp. 52--63. Springer Berlin Heidelberg, Berlin,
  Heidelberg (2005)

\bibitem{preibisch:09}
Preibisch, S., Saalfeld, S., Tomancak, P.: {Globally optimal stitching of tiled
  3D microscopic image acquisitions}.
\newblock Bioinformatics \textbf{25}(11), 1463--1465 (2009)

\bibitem{Press2001}
Press, W.A., Olshausen, B.A., Essen, D.C.V.: {A graphical anatomical database
  of neural connectivity}.
\newblock Philosophical Transactions of the Royal Society \textbf{356}(1412),
  1147--1157 (2001)

\bibitem{prckovska_fused_2011}
Pr\v{c}kovska, V., Peeters, T.H., van Almsick, M., ter Haar~Romeny, B.,
  Vilanova~i Bartroli, A.: Fused {DTI/HARDI} visualization.
\newblock {IEEE} Transactions on Visualization and Computer Graphics
  \textbf{17}(10), 1407--1419 (2011)

\bibitem{vazquez:09}
Reina, A.V., Miller, E., Pfister, H.: Multiphase geometric couplings for the
  segmentation of neural processes.
\newblock IEEE Conference on Computer Vision and Pattern Recognition pp. 1--8
  (2009)

\bibitem{roberts:11}
Roberts, M., Jeong, W.K., V, A., Unger, M.: {Neural Process Reconstruction from
  Sparse User Scribbles}.
\newblock In: Medical Image Computing and Computer Assisted Intervention, pp.
  1--8 (2011)

\bibitem{Rubinov2009}
Rubinov, M., Sporns, O.: {{C}omplex network measures of brain connectivity:
  {Uses} and interpretations}.
\newblock Neuroimage \textbf{52}, 1059--1069 (2010)

\bibitem{rybak2010digital}
Rybak, J., Kuss, A., Lamecker, H., Zachow, S., Hege, H., Lienhard, M., Singer,
  J., Neubert, K., Menzel, R.: The digital bee brain: integrating and managing
  neurons in a common 3d reference system.
\newblock Frontiers in systems neuroscience \textbf{4} (2010)

\bibitem{salvador05cerebralcortex}
Salvador, R., Suckling, J., Coleman, M.R., Pickard John, D., Menon, D.,
  Bullmore, E.: Neurophysiological architecture of functional magnetic
  resonance images of human brain.
\newblock Cereb Cortex \textbf{15}, 1332--1342 (2005)

\bibitem{schomer10:_nieder_elect}
Schomer, D.L., Lopes~da Silva, F.: {Niedermeyer's Electroencephalography: Basic
  Principles, Clinical Applications, and Related Fields}.
\newblock Wolters Kluwer/Lippincott Williams \& Wilkins (2010)

\bibitem{schultz_topological_2007}
Schultz, T., Theisel, H., Seidel, H.P.: Topological visualization of brain
  diffusion {MRI} data.
\newblock {IEEE} Transactions on Visualization and Computer Graphics
  \textbf{13}(6), 1496--1503 (2007)

\bibitem{seung:11}
Seung, S.: Connectome.
\newblock Houghton Mifflin Harcourt (2011).
\newblock In press

\bibitem{sharan06:_model}
Sharan, R., Ideker, T.: Modeling cellular machinery through biological network
  comparison.
\newblock Nature Biotechnology \textbf{24}(4), 427--433 (2006)

\bibitem{Sherbondy2005}
Sherbondy, A., Akers, D., Mackenzie, R., Dougherty, R., Wandell, B.: {Exploring
  connectivity of the brain's white matter with dynamic queries}.
\newblock IEEE Transactions on Visualization and Computer Graphics
  \textbf{11}(4), 419--430 (2005)

\bibitem{shu09:_alter_anatom_networ_in_early}
Shu, N., Liu, Y., Li, J., Li, Y., Yu, C., Jiang, T.: Altered anatomical network
  in early blindness revealed by diffusion tensor tractography.
\newblock PLoS One \textbf{4}(9), e7228 (2009)

\bibitem{Sporns2010}
Sporns, O.: {Networks of the brain}.
\newblock MIT Press (2010)

\bibitem{sporns:05}
Sporns, O., Tononi, G., K\"{o}tter, R.: {The Human Connectome: A Structural
  Description of the Human Brain}.
\newblock PLoS Computational Biology \textbf{1}(4), e42 (2005)

\bibitem{sporns04}
Sporns, O., Zwi, J.: The small world of the cerebral cortex.
\newblock Neuroinformatics \textbf{2}, 145--162 (2004)

\bibitem{stam07:_graph_theor_analy_of_compl}
Stam, C.J., Reijneveld, J.C.: Graph theoretical analysis of complex networks in
  the brain.
\newblock Nonlinear Biomedical Physics \textbf{1} (2007)

\bibitem{straehle:11}
Straehle, C., K\"{o}the, U., Knott, G., Hamprecht, F.: {Carving: Scalable
  Interactive Segmentation of Neural Volume Electron Microscopy Images}.
\newblock In: MICCAI, pp. 657--664 (2011)

\bibitem{tuch_qball_2004}
Tuch, D.S.: Q‐ball imaging.
\newblock Magnetic Resonance in Medicine \textbf{52}(6), 1358--1372 (2004)

\bibitem{tuch_high_2002}
Tuch, D.S., Reese, T.G., Wiegell, M.R., Makris, N., Belliveau, J.W., Wedeen,
  V.J.: High angular resolution diffusion imaging reveals intravoxel white
  matter fiber heterogeneity.
\newblock Magnetic Resonance in Medicine \textbf{48}(4), 577--582 (2002)

\bibitem{turetken:11}
T\"{u}retken, E., Gonz\'{a}lez, G., Blum, C., Fua, P.: {Automated
  Reconstruction of Dendritic and Axonal Trees by Global Optimization with
  Geometric Priors}.
\newblock Neuroinformatics \textbf{9}(2), 279--302 (2011)

\bibitem{valiant:06}
Valiant, L.G.: {A quantitative theory of neural computation}.
\newblock Biological Cybernetics \textbf{95}(3), 205--211 (2006)

\bibitem{van_dixhoorn_visual_2010}
Van~Dixhoorn, A., Vissers, B., Ferrarini, L., Milles, J., Botha, C.P.: Visual
  analysis of integrated resting state functional brain connectivity and
  anatomy.
\newblock In: Proc. Eurographics Workshop on Visual Computing for Biomedicine,
  VCBM, Leipzig, Germany, pp. 57--–64 (2010)

\bibitem{varshney:11}
Varshney, L.R., Chen, B.L., Paniagua, E., Hall, D.H., Chklovskii, D.B.:
  {Structural Properties of the Caenorhabditis elegans Neuronal Network}.
\newblock PLoS Computational Biology \textbf{7}(2), 21 (2011)

\bibitem{vazquez:11}
Vazquez-Reina, A., Pfister, H., Miller, E.L.: {Segmentation Fusion for
  Connectomics}.
\newblock International Conference on Computer Vision pp. 1--8 (2011)

\bibitem{vilanova_introduction_2005}
Vilanova, A., Zhang, S., Kindlmann, G., Laidlaw, D.H.: An introduction to
  visualization of diffusion tensor imaging and its applications.
\newblock In: Visualization and Image Processing of Tensor Fields.
  {Springer-Verlag} (2005)

\bibitem{vitaladevuni:10}
Vitaladevuni, S.N.: Co-clustering of image segments using convex optimization
  applied to em neuronal reconstruction.
\newblock IEEE Conference on Computer Vision and Pattern Recognition pp.
  2203--2210 (2010)

\bibitem{Walter2010}
Walter, T., Shattuck, D.W., Baldock, R., Bastin, M.E., Carpenter, A.E., Duce,
  S., Ellenberg, J., Fraser, A., Hamilton, N., Pieper, S., Ragan, M.a.,
  Schneider, J.E., Tomancak, P., H\'{e}rich\'{e}, J.K.: {Visualization of image
  data from cells to organisms}.
\newblock Nature Methods \textbf{7}(3s), S26--S41 (2010)

\bibitem{Wan2009}
Wan, Y., Otsuna, H., Chien, C.B., Hansen, C.: {An interactive visualization
  tool for multi-channel confocal microscopy data in neurobiology research}.
\newblock IEEE Transactions on Visualization and Computer Graphics
  \textbf{15}(6), 1489--1496 (2009)

\bibitem{White1986}
White, J.G., Southgate, E., Thomson, J.N., Brenner, S.: {The Structure of the
  Nervous System of the Nematode Caenorhabditis elegans}.
\newblock Philosophical Transactions of the Royal Society B Biological Sciences
  \textbf{314}(1165), 1--340 (1986)

\bibitem{worsley_comparing_2005}
Worsley, K.J., Chen, J., Lerch, J., Evans, A.C.: Comparing functional
  connectivity via thresholding correlations and singular value decomposition.
\newblock Philosophical Transactions of the Royal Society B: Biological
  Sciences \textbf{360}(1457), 913 --920 (2005)

\bibitem{zalesky10:_networ}
Zalesky, A., Fornito, A., Bullmore, E.T.: Network-based statistic: identifying
  differences in brain networks.
\newblock Neuroimage \textbf{53}(4), 1197--1207 (2010)

\bibitem{song_zhang_visualizing_2003}
Zhang, S., Demiralp, C., Laidlaw, D.H.: Visualizing diffusion tensor {MR}
  images using streamtubes and streamsurfaces.
\newblock {IEEE} Transactions on Visualization and Computer Graphics
  \textbf{9}(4), 454-- 462 (2003)

\end{thebibliography}
